\documentclass[11pt]{article}
\usepackage[a4paper]{geometry}

\usepackage{datetime}
\usepackage{lastpage}
\usepackage{fancyhdr}

\newenvironment{multicols}[1]{}{} 

\usepackage{textcase}

\usepackage{titlesec}
\titleformat{\section}
{\sffamily\normalsize\filcenter}
{\thesection.}{.5em}{\MakeTextUppercase}

\titleformat{\subsection}
{\sffamily\normalsize}
{\thesubsection}{.5em}{}
\titleformat{\subsubsection}
{\sffamily\normalsize}
{\thesubsubsection}{.5em}{}
\titleformat{\paragraph}[runin]
{\normalsize}
{}{.5em}{}[.]
\titlespacing*{\paragraph}{\parindent}{*1}{*1}

\renewenvironment{table}[1][]{}{}
\renewenvironment{figure}[1][]{}{}

\usepackage{xspace, booktabs}
\usepackage{qtimes}
\usepackage{tgheros}

\usepackage{natbib}
\bibpunct{(}{)}{;}{a}{}{,}
\def\bibfont{\footnotesize}

\usepackage{caption} 
\usepackage{ifthen}
\newif\ifblind\blindfalse

\newcommand{\authorfootnote}[1]{\def\TheAuthorFootnote{#1}}

\renewenvironment{abstract}{\hfil\begin{minipage}{.85\textwidth}\noindent\hrulefill\par\noindent\small}
  {\noindent\hrulefill\end{minipage}\hfil}

\makeatletter
\newcount\auth@rno
\renewcommand{\author}[2][]{
 \advance\auth@rno by 1 \relax%
 \expandafter\def\csname anAuth@r\number\auth@rno\endcsname{#2}
 \def\short@uthor{#1}%
}

\newcommand{\PrintAuthor}[1]{{\csname anAuth@r#1\endcsname}}
\newcount\zaehl
\newcount\endauth

\newcommand{\PrintAuthors}
{ \ifnum\auth@rno<1{No Author? }\fi
  \zaehl=1
  \endauth = \auth@rno
  \advance\endauth by -1 \relax%
  \loop\PrintAuthor{\number\zaehl}%
  \ifnum \zaehl< \auth@rno
   {\ifnum \zaehl<\endauth,\ \else,\ and\ \fi} 
    \advance\zaehl by 1\relax %
  \repeat
}

\makeatother

\makeatletter
\renewcommand{\title}[2][]{\def\@title{#2}\def\@shorttitle{#1}}
\makeatother
\newcommand{\keywords}[1]{\medskip\par\noindent KEY WORDS: #1\par}

\makeatletter

\def\maketitle{%
\begin{center}
\par{\bfseries\sffamily\flushright\LARGE\@title}%
\par{\sffamily\flushleft\large \PrintAuthors}\medskip
\par{\sffamily\today, \currenttime}\medskip
\end{center}
\let\thefootnote\relax\footnotetext{\TheAuthorFootnote
\ifx\TheAcknowledgement\undefined\relax\else\par\TheAcknowledgement\fi}
\par}

\makeatother
\renewcommand{\footnotemark}{}

\providecommand{\email}[1]{(E-mail: \textit{#1})}
\usepackage[english]{babel}

\usepackage{amsmath}

\usepackage[T1]{fontenc} 
\usepackage[utf8]{inputenc} 

\usepackage{amssymb}

\usepackage{graphicx}

\usepackage{array}     

\usepackage[breaklinks]{hyperref}  

\usepackage[footnote, marginclue]{fixme}

\title[Rank correction to multiple testing]
{Multiple Monte Carlo Testing with Applications in Spatial Point Processes}

\begin{document}

\author{Tom\' a\v s Mrkvi\v cka}
\author{Mari Myllym{\"a}ki}
\author[Mrkvi\v cka,  Myllym{\"a}ki, Hahn]{Ute Hahn}

\authorfootnote{
Tom\' a\v s Mrkvi\v cka is Associate Professor, Department of Applied Mathematics and Informatics, Faculty of Economics, University of South Bohemia, \v{C}esk\'e Bud\v{e}jovice, Czech Republic \email{mrkvicka.toma@gmail.com}. Mari Myllym{\"a}ki is researcher, Natural Resources Institute, Espoo, Finland \email{mari.myllymaki@luke.fi}.
Ute Hahn is Associate Professor, Centre for Stochastic Geometry and Advanced Bioimaging, Department of Mathematics, University of Aarhus, Aarhus, Denmark \email{ute@imf.au.dk}.
}

\maketitle

\begin{abstract}
The rank envelope test (Myllym{\"a}ki et al., Global envelope tests for spatial processes,  arXiv:1307.0239 [stat.ME]) is proposed as a solution to multiple testing problem for Monte Carlo tests. Three different situations are recognized: 1) a few univariate Monte Carlo tests, 2) a Monte Carlo test with a function as the test statistic, 3) several Monte Carlo tests with functions as test statistics. The rank test has correct (global) type I error in each case and it is accompanied with a $p$-value and with a graphical interpretation which shows which subtest or which distances of the used test function(s) lead to the rejection at the prescribed significance level of the test. Examples of null hypothesis from point process and random set statistics are used to demonstrate the strength of the rank envelope test. The examples include goodness-of-fit test with several test functions, goodness-of-fit test for one group of point patterns, comparison of several groups of point patterns, test of dependence of components in a multi-type point pattern, and test of Boolean assumption for random closed sets.

\end{abstract}

\begin{multicols}{2}
\section{Introduction}

Nowadays, Monte Carlo tests are used in many applications. In particular, these tests are used in fields where no analytical results are usually available. One such field is spatial statistics. In our work, we concentrate mainly on a subfield of spatial statistics, spatial point processes.
In a Monte Carlo test, a test statistic $\mathbf T$ is chosen and the statistic estimated from the data ($\mathbf T_1$) is compared to $s$ simulated statistics obtained from simulations under the null hypothesis ($\mathbf T_2, \dots, \mathbf T_{s+1}$). If the data statistic $\mathbf T_1$ takes an extreme rank among all the statistics, the null hypothesis is rejected. This kind of Monte Carlo test was introduced by \citet{Barnard1963} and popularized for spatial point patterns by \citet{BesagDiggle1977}. Throughout this paper we will consider this type of Monte Carlo test only.

The chosen test statistic $\mathbf T$ can be univariate or multivariate. This paper considers the multivariate case where the components of the vector are generally dependent. In this case, the usual way to perform a test is to transform $\mathbf T$ to a univariate case, in order to avoid the multiple testing problem which arises since one wants to base the test on all $m>1$ components of $\mathbf T$.
Such a solution is called deviation test in point process statistics. In these deviation tests \citep[see e.g.][]{IllianEtal2008, MyllymakiEtal2015a}, the (scaled) maximum or integral measure of all components is used as a transformation measure. In this work, we consider another solution to overcome the problem of multiple testing, which is the extreme rank ordering \citep{MyllymakiEtal2015b}. This ordering gives exactly the same weight to every component of $\mathbf T$ and it provides graphical visualization for all components, which is seen by practitioners as a great advantage. The rank test is the Monte Carlo test based on the extreme rank ordering.

Another solution to multiple testing problem is to use Bonferroni-type corrections \citep{Simes1986, Hommel1988, Hochberg1988, Rom1990}. Such methods are rather conservative especially in the case where large number of dependent tests are considered. Further, \citet{BenjaminiHochberg1995} introduced a method based on controlling false discovery rate, which weakly controls the global type I error. On the other hand, the rank test considered in this paper is exact (in the sense of correct type I error under a simple null hypothesis) for any number of Monte Carlo tests either dependent or independent and any number of used simulations $s$.

The rank envelope test was first introduced by \citet{MyllymakiEtal2015b} for the case where $\mathbf T$ is a test function, which is in practice discretized to a high dimensional test vector. \citet{MyllymakiEtal2015b} used the rank test for goodness-of-fit testing of point process models. This test provides both an exact $p$-value and graphical visualization. The graphical visualization is given by the $100\%(1-\alpha)$ simultaneous envelope which has the intuitive meaning: If the data test vector is outside the simultaneous envelope (at least for one component of the test vector) the null hypothesis is rejected at the prescribed level $\alpha$. In this paper, we generalize this idea for a general test vector of any dimension. Especially, we show how this procedure can be used 1) for a test vector with low number of components, 2) for a test vector with many highly correlated components \citep[suggested in][]{MyllymakiEtal2015b} and 3) for a test vector with almost independent blocks with high inner correlation. The last case covers a rank test based on several test functions, i.e.\ combining several rank tests where each rank test is performed on a different test function. It also allows to make a post-hoc comparison of such a combined rank test. We show several possible usage of the rank test in these situations by examples taken from spatial statistics.

First of all we show, in Section 4, the use of low dimensional test vector as a solution of multiple testing problem for goodness-of-fit test of Boolean model.

In Section 5 we use the rank test to solve the multiple testing problem of several goodness-of-fit tests performed with different test functions on the same data. Usually it is not known in advance which test function is sensitive to reveal deviations from the given null hypothesis. If one wants to use several test functions, then the combined rank test can be employed to obtain one common $p$-value for the combined test. The graphical visualization is possible and the simultaneous envelope is given jointly for all test functions (on the global level $\alpha$). We also perform a simulation study to explore how the power is affected by the use of several test functions in comparison to using only one. 

In Section 6, we consider a goodness-of-fit test for replicated point pattern data and show how the combined rank test can be used to overcome the multiple testing problem. The combined rank test provides one common $p$-value and also identifies which of the point patterns is the reason for the (potential) rejection of the null model. Testing and identification are made on the global level $\alpha$.

In Section 7, we discuss the problem of comparing several groups of point patterns by combined rank test, which leads to a kind of nonparametric functional ANOVA. An advantage of the combined rank test is that it immediately provides graphical post-hoc comparison, which is done on the global level $\alpha$.

In Section 8, we address the problem of dependence of components of a multi-type point pattern with more than two types.

The rank test and also its graphical visualization is first explained in detail in Section 2. In Section 3, the number of simulations needed to perform the rank test is discussed. Section 9 is devoted to further 	discussion.

The proposed method is provided in an R library \emph{spptest}, which can be obtained at \\
\texttt{https://github.com/myllym/spptest}.

\newcommand{\TT}{\mathbf{T}}
\newcommand{\lo}{\mathrm{low}}
\newcommand{\up}{\mathrm{upp}}
\providecommand{\Pr}{\mathrm{Pr}}
\providecommand{\1}{\mathbf{1}}
\section{Multivariate Monte Carlo tests based on pointwise ranks}

The idea of the multiple Monte Carlo testing considered in this paper is based on the rank envelope test introduced in \cite{MyllymakiEtal2015b}. In the mentioned paper the rank envelope test was considered in detail for a functional test statistic, which is typically an estimator of a (well-known) summary function. Here, in the present paper, we extend this idea into general multivariate vector of the form
\[
\TT = \big(T_1,\dots,T_d).
\]

This extension enables us to consider various test hypothesis, which are not covered in the original case of \cite{MyllymakiEtal2015b}.
In Section \ref{sec:lowdimensional}, $\TT$ consists of only a few measurements of intrinsic volumes on a random closed set, in Section \ref{sec:combinedfunctions}, values of several different summary functions estimated on a point pattern are combined into one vector, in Sections \ref{sec:gof-several-patterns} and \ref{sec:ANOVA}, the vector consists of estimates from the same summary function on several patterns. Finally, in Section \ref{sec:multitype}, the vector consists of estimates of different summary characteristics of a multivariate point pattern.

Further, we define also a one sided test, whereas in the previous work only two sided tests were considered. Although, the extensions considered in this Section are rather straightforward, we briefly define these extensions.

\subsection{Rank envelope test}

Let $\TT_1$ be the observed statistic, and $\TT_2, \dots, \TT_{s+1}$ be a sample of $s$ realizations of $\TT$ under the null hypothesis.
The rank envelope test \citep{MyllymakiEtal2015b} with level $\alpha$ constructs a set $\{\TT_{\lo}^{(\alpha)}, \TT_{\up}^{(\alpha)}\}$ of envelope vectors such that, under the simple null hypothesis, the probability that $\TT_1=(T_{11}, \dots, T_{1d})$ falls outside this envelope in any of the $d$ points is less or equal to $\alpha$,
\begin{equation*}
  \Pr\big(T_{1j}\notin[T_{\lo\,j}^{(\alpha)},T_{\up\,j}^{(\alpha)}]\ \text{for any $j$}\,\big|\, H_0\big)\leq \alpha
\end{equation*}
and
 the probability that $\TT_1$ falls outside this envelope or touches it in any of the $d$ points is greater than $\alpha$,
\begin{equation*}
  \Pr\big(T_{1j}\notin(T_{\lo\,j}^{(\alpha)},T_{\up\,j}^{(\alpha)})\ \text{for any $j$}\,\big|\, H_0\big) > \alpha
\end{equation*}


In goodness of fit tests, the realizations $\TT_2, \dots, \TT_{s+1}$ are independent (or at least exchangeable), and generated by simulation. There is also a possibility to generate the realizations by permuting samples of the observed data. Such tests are often called permutation tests. Simulation based tests are dealt with in Sections \ref{sec:lowdimensional}, \ref{sec:combinedfunctions}, \ref{sec:gof-several-patterns}, and \ref{sec:multitype}, while Section \ref{sec:ANOVA} uses permutation.

\subsubsection{Calculation of $p$-values}

The test is easiest to understand from perspective of the associated $p$-values. According to the framework of Barnard's Monte Carlo test or Pitman's permutation test, $p$-values are obtained by assigning an extreme rank $R_i$ to each of the vectors $\TT_i$, such that the lowest ranks correspond to the most extreme values of the statistic. The conservative and liberal $p$-values are then given as
  \begin{equation}\label{eq:p-value_globalranktest}
   p_+ =  \sum_{i=1}^{s+1} \1 (R_i \leq R_1)  \big/ (s+1), \quad  p_- =  \sum_{i=1}^{s+1} \1 (R_i < R_1)  \big/ (s+1).
  \end{equation}
The extreme rank $R_i$ of the vector $\TT_i$ is the minimum of the pointwise ranks $R_{ij}, j=1, \ldots , d$ of its elements of $T_{ij}$ among the corresponding elements $T_{1j}, T_{2j}, \dots,T_{(s+1)j}$ in all $s+1$ vectors,
  \begin{equation}\label{eq:minglobalrank}
  R_i=\min_j R_{ij}.
  \end{equation}
How the element wise ranks are determined, depends on whether a one sided or a two sided test is to be performed.
Let $r_{1j}, r_{2j}, \dots, r_{(s+1)j}$ be the raw ranks of $T_{1j}, T_{2j}, \dots, T_{(s+1)j}$, such that the smallest $T_{ij}$ has rank 1. In the case of ties, the raw ranks are averaged. The resulting pointwise ranks are calculated as
\begin{equation}
  R_{ij}=\begin{cases}
    r_{ij}, &\text{one-sided test, small $T$ is considered extreme}\\
    s+1-r_{ij}, &\text{one-sided test, large $T$ is considered extreme}\\
   \min(r_{ij}, s+1-r_{ij}), &\text{two-sided test}.
  \end{cases}
\end{equation}

\subsubsection{The graphical envelope test}

For the graphical version of the test, an appropriate low rank $R_{(\alpha)}$ is determined as shown below, and the envelope is constructed as the hull of those ``less extreme'' vectors $\TT_i$ that have rank $R_i \geq R_{(\alpha)}$. Let $I_\alpha = \{i\in 1,\dots, s+1: R_i \geq R_{(\alpha)}\}$ be the index set of vectors, and define
\begin{equation}\label{rank_envelopes}
  T_{\lo\,j}^{(\alpha)}= \min_{i\in I_\alpha}T_{ij}, \quad   T_{\up\,j}^{(\alpha)}= \max_{i\in I_\alpha}T_{ij}
\end{equation}
for the two sided test. For one-sided tests, let $T_{\lo\,j}^{(\alpha)}=-\infty$ or $T_{\up\, j}^{(\alpha)}=\infty$, respectively.
By choosing  $R_{(\alpha)}$ as the smallest value in $\{R_1,\dots,R_{s+1}\}$ for which
\begin{equation}\label{eq:Ralpha}
  \sum_{i=1}^{s+1} \1 \left(R_i \leq R_{(\alpha)}\right) \geq \alpha(s+1),
\end{equation}
we get the following interpretation.

If the observed vector leaves this envelope in some point i.e. $R_1 < R_{(\alpha)}$, which is equivalent to $p_+ \leq \alpha$, the null hypothesis is rejected. If the observed vector is completely inside this envelope i.e. $R_1 > R_{(\alpha)}$, which is equivalent to $p_- > \alpha$, the null hypothesis is not rejected. If the observed vector coincides in some point with the border of this envelope, i.e. $R_1 = R_{(\alpha)}$, which is equivalent to $p_- \leq \alpha < p_+ $, the rejection of the null hypothesis remains undecided.

The above interpretation is a direct consequence of Theorem 4.2 in \citet{MyllymakiEtal2015b} for the two sided rank envelope test. The proof of this theorem can be done in the same way also for the one sided rank (envelope) test.

\subsection{The problem of ties, and $p$-intervals}
If the extreme ranks $R_i$ were almost surely different, the $p$-values of the global envelope test would take the values $1/(s+1), 2/(s+1),\dots, 1$ with equal probability under the null hypothesis. However, due to the construction as vector wise minimum of pointwise ranks \eqref{eq:minglobalrank}, ties occur very often. In a one sided test with $d$-variate vectors, up to $d$ out of the $s+1$ vectors could take rank 1. Therefore instead of a single $p$-value, \citet{MyllymakiEtal2015b} suggest to accompany the test with a $p$-interval  $(p_-, p_+ ]$.
The length of the $p$-interval,
\[
p_+ - p_- = \frac{1}{s+1}\sum_{i=1}^{s+1} \1 (R_i = R_1),
\]
determines the "grey" zone of the test. It was shown in \citet{MyllymakiEtal2015b} that this length is of order $s^{-1}$. However, it also depends on the correlation structure of the multivariate vectors. In Section \ref{sec:numsimul}, we investigate the needed number of simulations $s$ with respect to the type of correlation structure of the multivariate vector.

\subsection{Breaking ties by extreme rank count ordering}
To resolve the problem of  "grey" zone of the test, \citet{MyllymakiEtal2015b} defined extreme rank count ordering which refines the extreme rank ordering in order to minimize the possibility of ties. We briefly rephrase the definition in the multivariate vector case.

Consider the vectors of pointwise ordered ranks $\mathbf{R}_i=(R_{i[1]}, R_{i[2]}, \dots , R_{i[d]})$, where \\ $\{R_{i[1]},  \dots , R_{i[d]}\}=\{R_{i1}, \dots, R_{id}\}$ and $R_{i[j]} \leq R_{i[j^\prime]}$ whenever $j \leq j^\prime$.

The extreme rank given in \eqref{eq:minglobalrank} corresponds to $R_i=R_{i[1]}$. It was suggested to replace this by the rank under extreme rank count ordering of the vectors $\mathbf{R}_i$, namely to consider ordering based on
\begin{equation}
\label{eq:lexicalrank}
   R_i^{\text{erc}} = \sum_{i^\prime=1}^{s+1} \1(\mathbf{R}_{i^\prime} \prec \mathbf{R}_i)
\end{equation}
where
\[
\mathbf{R}_i \prec \mathbf{R}_{i^\prime} \quad \Longleftrightarrow\quad
  \exists\, n\leq d: R_{i[j]} = R_{i^\prime[j]} \forall j < n,\  R_{i[n]} < R_{i^\prime[n]}.
\]

\subsection{The type I error}
The possibility of ties in the extreme rank count ordering is rather small, therefore the rank (envelope) test with extreme rank count ordering as a solution for the ties has the exact type I error under simple null hypotheses in practice.

In the case of composite null hypothesis, where some parameters of the null model have to be estimated, the test is usually conservative. The amount of the conservativeness depends on the correlation of the estimating and testing functions and on the precision of the  estimation procedure.  However, the test can be instead liberal, if the estimation procedure is biased as it is shown in Section 4. \citet{MyllymakiEtal2015b} showed the possibility of applying the procedure described in \cite{DaoGenton2014} on rank envelope test. This adjusted rank test corrects the type I error of the test under a composite hypothesis, but it is rather time consuming procedure, because it requires $s^2$ simulations. 
A composite null hypothesis is tested in Section 4, where the adjusted rank test is applied. The composite null hypothesis is also tested in a part of the simulation study of Section 5. There, we use only pure rank test due to the time constraints. Note that this simplification does not influence the conclusions made from the simulation study.

\section{Appropriate number of simulations}\label{sec:numsimul}

Note first that the rank test with extreme rank count has exact type I error under the simple null hypothesis with whatever number of simulations. Only the precision of the graphical interpretation, which is given by a width of $p$-interval, can be unsatisfactory. In this section, we give some recommendations for the number of simulations $s$ for the common choice of the significant level, $\alpha=0.05$. For this significant level we would like the width of $p$-interval to be 0.01 at maximum.

\subsection{Number of simulations for a low dimensional test vector}

Having a test vector of dimension $d$, the maximal width of the $p$-interval is simply
\begin{equation}\label{simple_ns}
2 d / (s+1)
\end{equation}
for a two-sided rank test and 
\begin{equation}\label{simple_nsone}
d / (s+1)
\end{equation}
for a one-sided rank test.

Usually, there is low correlation between components of the test vector when $d$ is small and, therefore, the above formulas can be used to determine the appropriate number of simulation in this case. In the case of a high dimensional test vector, the formula \eqref{simple_ns} or \eqref{simple_nsone} gives an upper limit for the width. However, the choice of $s$ based on this upper limit would be too time consuming.

\subsection{Number of simulations for a test function}

A test function is in practice a high-dimensional test vector with high correlation between the components. (In our studies, we have used $d = 500$.) Due to the high correlation the width of the $p$-interval is much smaller than the upper limit given by \eqref{simple_ns} or \eqref{simple_nsone}. In our previous study \citep{MyllymakiEtal2015b}, where only this case was studied, it was recommended to use at least 2500 simulations when testing at the significance level 0.05.

\subsection{Number of simulations for a combination of several test functions}

The rank test for combination of several test functions can be seen as a two stage procedure. The first step is to compute the extreme rank ordering for each sub-test, i.e.\ for each test function separately. The global extreme rank $R_i$ is then the minimum of the extreme ranks from the sub-tests. Thus the second step can be seen as a one-sided rank test performed on the extreme ranks computed in sub-tests. Since, generally in different sub-tests, different simulations contibute to the most extreme rank, the recommended number of simulations for a combination of $k$ test functions is $k$ * 2500.

\subsection{Number of simulation for the rank test with extreme rank count as a solution for ties}

The extreme rank count is practically a continuous test statistic. Therefore the probability of having a tie in extreme rank counts is very small and can be disregarded. This means that it is possible to use extreme rank count ordering with less simulations than the extreme rank ordering, but then the graphical envelope interpretation may be lost, because the data function may coincide a boundary of the envelope.

We remark here that classically in a Monte Carlo test the $p$-value is estimated from the given simulations of the test statistics ($u_i$ in the deviation test, $R_i$ in the rank test). The standard deviation of such a point estimate decreases with the square root of the number of simulations performed. \cite{LoosmoreFord2006} recommended to use at least 999 simulations to reduce this uncertainty.

\newcommand{\rd}{{\mathbb R}^d}
\newcommand{\ep}{\varepsilon}
\newcommand{\dist}{{\rm dist}\,}
\newcommand{\R}{{\mathbb R}}

\section{Test with use of low dimensional random vector}\label{sec:lowdimensional}

\begin{center}
\begin{figure}[htbp]
\centerline{\includegraphics[width=4.5cm,angle=0]{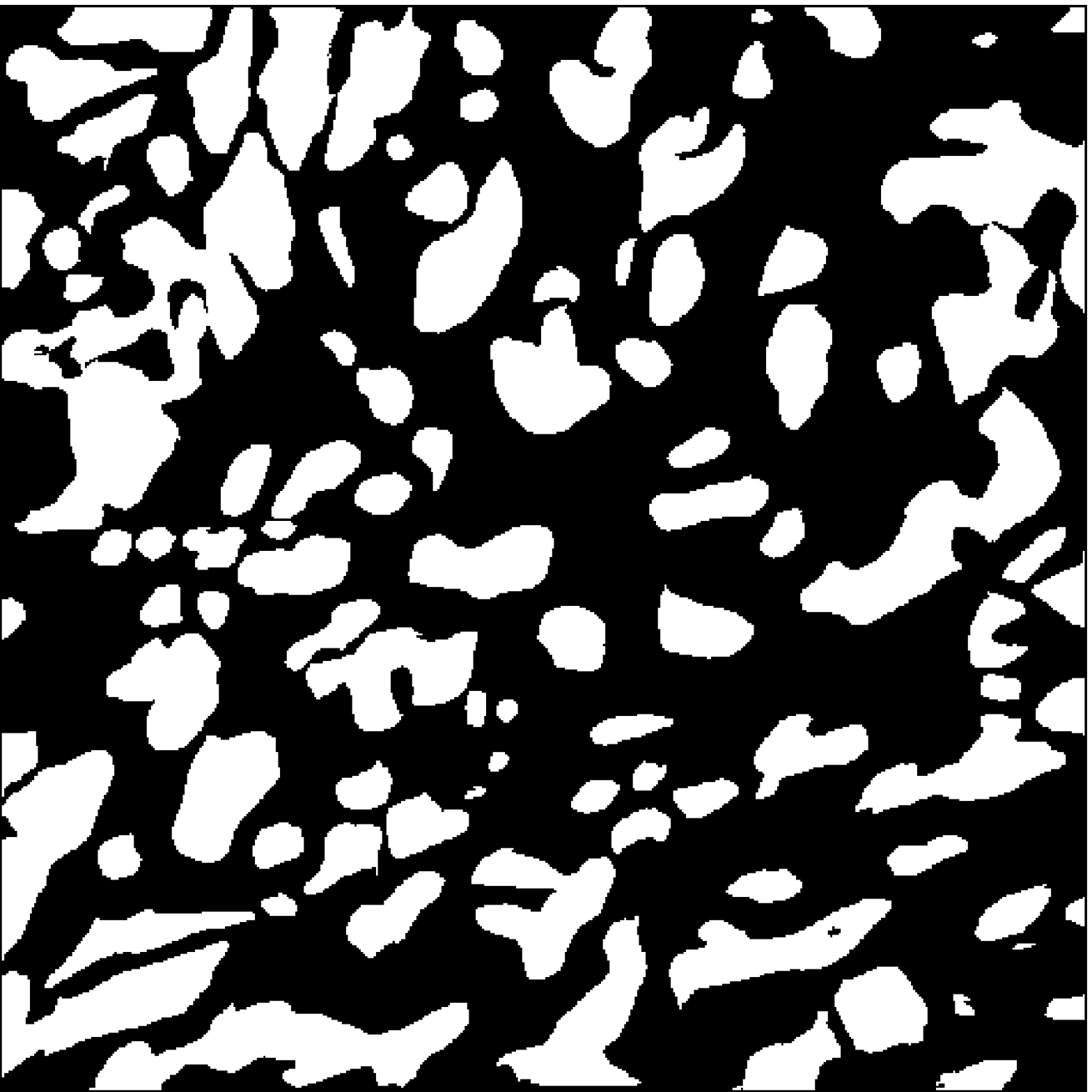} \hspace{0.5cm}
\includegraphics[scale=0.25]{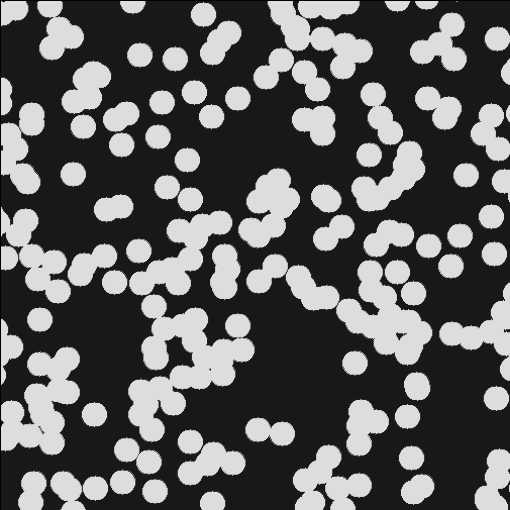} }
\captionof{figure}{\label{cancer}Left: A binary image of mammary cancer tissue with resolution of 512 $\times $ 512 pixels. Right: A realization of the fitted Boolean model of disks with lognormal distribution of disk radii.}
\end{figure}
\end{center}

In this section, we demonstrate the rank test for combining several univariate Monte Carlo tests together.
As an example, we use 
the Boolean model of disks \citep[see e.g.][]{StoyanEtal1995, Mrkvicka2009}
as a null model for an image of mammary cancer tissue, see Figure \ref{cancer}, which is regarded as a random closed set. This data originate from a collection of 200 images studied in detail in \citet{MrkvickaMattfeldt2011},

We chose the distribution of disk radii in the Boolean model to be log normal and estimated the parameters of the  model by means of the contact distribution function \citep{Molchanov1995}.
A realization of the fitted process is shown in Figure \ref{cancer}. (The difference of the data and fit is mainly in the shape of sets, because the fitted model use only disks, therefore the chosen test statistics are not heavily dependent on the shape of sets.) Next, in order to conduct a test, simulations were generated from the fitted null model.

As test statistics we choose intrinsic volumes,
because they are the most important characteristics of random closed sets and
because they are not related to the characteristic used for estimation,
i.e.\ the contact distribution function.
In $l$-dimensional Euclidian space, the intrinsic volumes $V_0(K),\ldots ,V_l(K)$ of a convex body $K\subseteq {\mathbb R}^l$ are
determined by the Steiner formula $$V_l(K_\ep)=\sum_{k=0}^l
\ep^k\omega_k V_{l-k}(K),$$ where $V_l$ is the volume
($l$-dimensional Lebesgue measure), $K_\ep=\{ x\in\rd:\,\dist
(x,K)\leq\ep\}$ the (closed) $\ep$-parallel set to $K$ and
$\omega_k$ denotes the volume of the unit ball in $\R^k$.
(Under a different normalization, they are known as quermassintegrals or
Minkowski functionals.)
The intrinsic volumes can be extended additively to polyconvex sets
(sets from the convex ring), for details see \cite{Schneider1993}.
In the plane, ${V}_2(K)$ is the volume, ${V}_1(K)$
is one half of the circumference of the border $\partial K$ and
${V}_0(K)$ is the Euler number.

\begin{center}
\begin{figure}[htbp]
\centerline{\includegraphics[scale=0.35]{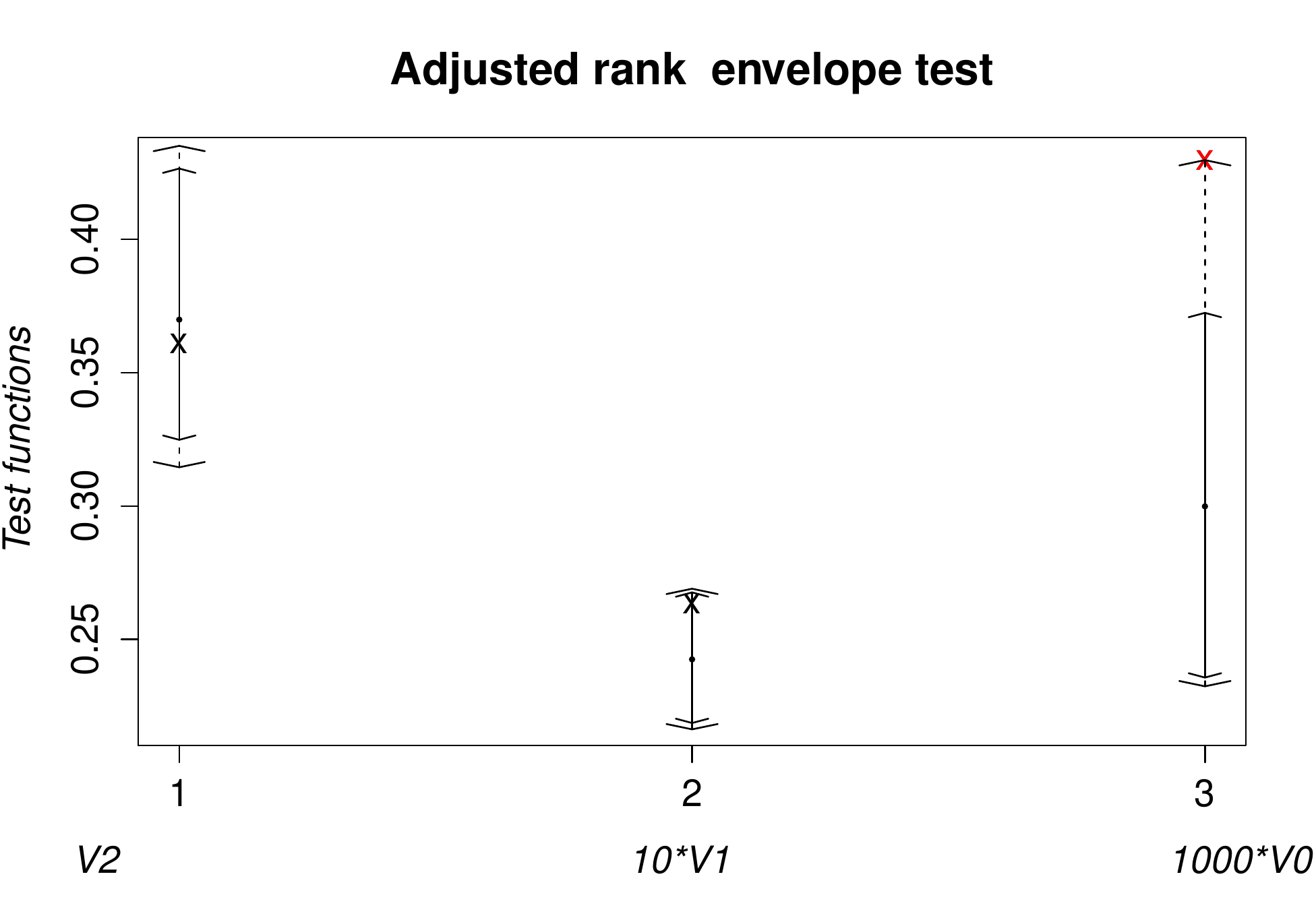}}
\captionof{figure}{\label{cancertest}The outer bounds show result of the adjusted rank envelope test with $299*300$ simulations of the null model where the test vector consists of all three intrinsic volumes, whereas the inner bounds show result of the rank envelope test with $299$ simulations. The crosses correspondes to the data intrinsic volumes.}
\end{figure}
\end{center}

We then performed the rank test with $s=299$ simulations
where the test vector is three-dimensional consisting of all three intrinsic volumes. The resulted $p$-interval is (0.003, 0.02) and the $p$-value based on the extreme rank count ordering is 0.013. Since we deal with composite hypothesis, we performed also adjusted rank test with $s(s+1)$ simulations \citep{MyllymakiEtal2015b}. 
The resulted graphical interpretation is shown in Figure \ref{cancertest}. The envelope for the adjusted test is wider than for the pure test, which refers to the estimation procedure not being the perfect one. (The histogram of $p$-values inside the adjusted test shows great preference for small $p$-values as well.) The resulted adjusted significance level $\alpha^\ast = 0.013$ which leads us to the rejection of the null hypothesis at the significance level 0.05. Furthermore, the graphical interpretation shows that the null hypothesis is rejected due to the Euler number which lies on the adjusted envelope, i.e.\ the data set has more isolated cells than the Boolean model.

\section{Goodness-of-fit test with many test functions simultaneously}\label{sec:combinedfunctions}

Deviation and envelope tests are the main tools for goodness-of-fit tests in spatial point process statistics \citep[see e.g.][]{IllianEtal2008, Diggle2013, MyllymakiEtal2015a, MyllymakiEtal2015b}. These tests are based on a test function $T(r)$ on a chosen interval $I$ of distances $r$.

For a test, one needs to choose $T(r)$. Previous experience and possible alternative hypotheses may suggest a test function to be used. However, often it is not known in advance which test function leads to a powerful test in the situation at hand, and one would like to employ several test functions, which however leads to multiple testing as such. 
The rank test can be used in this situation both for combining several deviation tests to one test and for combining several rank envelope tests \citep{MyllymakiEtal2015b} to one test. 

In the deviation test, the discrepancy between the (empirical and simulated) test functions $T_i(r)$, $i=1,\dots,s+1$, and the expectation of $T(r)$ under the null hypothesis $H_0$ are summarized in single values $u_i$, $i=1,\dots,s+1$, by a deviation measure, e.g.\ the (scaled) integrated discrepancy of all distances on $I$ or the (scaled) maximum discrepancy over the distances on $I$, see \citet{IllianEtal2008} and \citet{MyllymakiEtal2015a}.
If the data value $u_1$ obtains an extreme rank among all the $u_i$s, the test leads to rejection of $H_0$.
To combine several deviation tests to one test by means of the rank test, the test vector $\TT_i$ is taken to consist of the deviation measures $u_i^{1}, u_i^{2}, \dots, u_i^{d}$, where $d$ is the number of deviation tests, i.e.\ the number of test functions used. Thus, we are dealing with a low dimensional test vector in the rank test as in Section \ref{sec:lowdimensional}. This rank test is one-sided, since only large values of $u$ are typically considered significant.

There is also a graphical interpretation available for the classical \citep{Ripley1981} and scaled \citet{MyllymakiEtal2015b} maximum deviation measure test.
Unfortunately, we loose such graphical interpretation in combining several test functions together. We obtain only graphical interpretation for the combined rank test, telling which test function leads to the possible rejection of $H_0$.

In the combined rank envelope test, the test vector is taken to consist of all values of the first test function followed by all values of the second test function, etc.
Thus, the length of the test vector becomes $d \times K$, where $d$ stands for the number of
test functions and $K$ for the number of distances $r$ (in our simulation study below $K=500$).
We consider the same number of distances $r$ for each test function in order to ensure that each test function has the same importance in the global test.
The rank test is two-sided.

In combining several rank envelope tests, we have the graphical interpretation for each individual test, which is a great advantage of the combined rank envelope test in comparison to the combined deviation test.

\begin{center}
\begin{figure}[htbp]
\centerline{\includegraphics[scale=0.25]{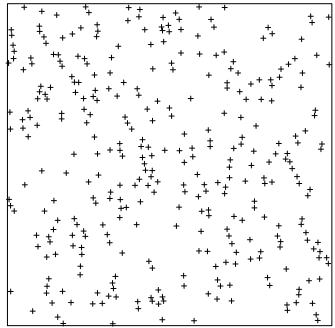}}
\captionof{figure}{\label{untreated} The first point pattern of intramembranous particles from control untreated group observed in a window 512 $\times$ 512 pixels.}
\end{figure}
\end{center}

In the study of \cite{SchladitzEtal2003}, the spatial structure of intramembranous particles was investigated separately by the $L$-, $F$-, $G$- and $J$-functions \citep[see e.g.][]{IllianEtal2008}. It was pointed out that some features of the spatial structure which are not visible by $L$-function can be visible by $G$-function. We use this data study to show advantages of our rank test.
Figure \ref{untreated} shows spatial locations of intramembranous particles taken from first sample of untreated group of the study of \cite{SchladitzEtal2003}.
We investigated a Gibbs hard core model \citep[i.e., the Strauss process where the interaction
parameter equals zero, see e.g.][]{StoyanEtal1995} as a null model for these locations of particles, as was done in \citet{MyllymakiEtal2015b} using one test function. 
We conditioned the model on the number of points, and fixed the only parameter, i.e. the hard core, to the minimum distance between two particles, i.e., 5.85 pixels in our sample, thus dealing with a simple hypothesis. 

\begin{center}
\begin{figure}[htbp]
\centerline{\includegraphics[width=\textwidth]{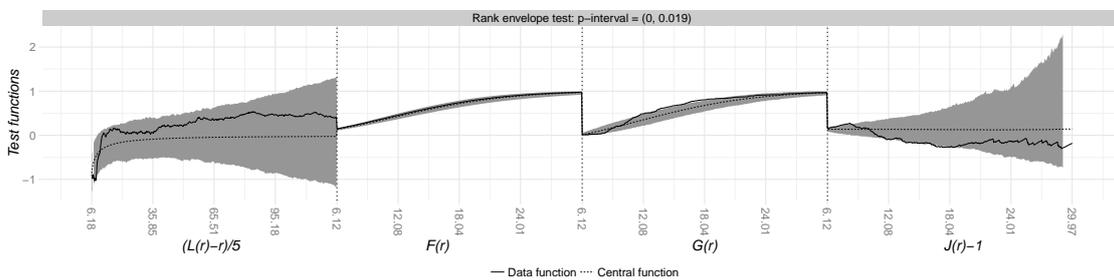} }
\captionof{figure}{\label{testLFGJ}The combined rank envelope test with $L(r)-r$, $F(r)$, $G(r)$ and $J(r)$ functions performed with $s=9999$ simulations of the null model. }
\end{figure}
\end{center}

The combined 95\% envelope is shown in Figure \ref{testLFGJ}. 
The test reveals deviation of the data from the null model for both small and medium values of $r$. But the deviation is proved by $L$ and $J$ functions for small $r$ while it is proved by $G$ and $J$ for medium $r$. The deviation for small $r$ is probably coused by the fact that the particles have variable size. While the deviation for medium $r$ is caused by clustering of particles for these distances.

\begin{center}
\begin{figure}[htbp]
\centerline{\includegraphics[width=0.5\textwidth]{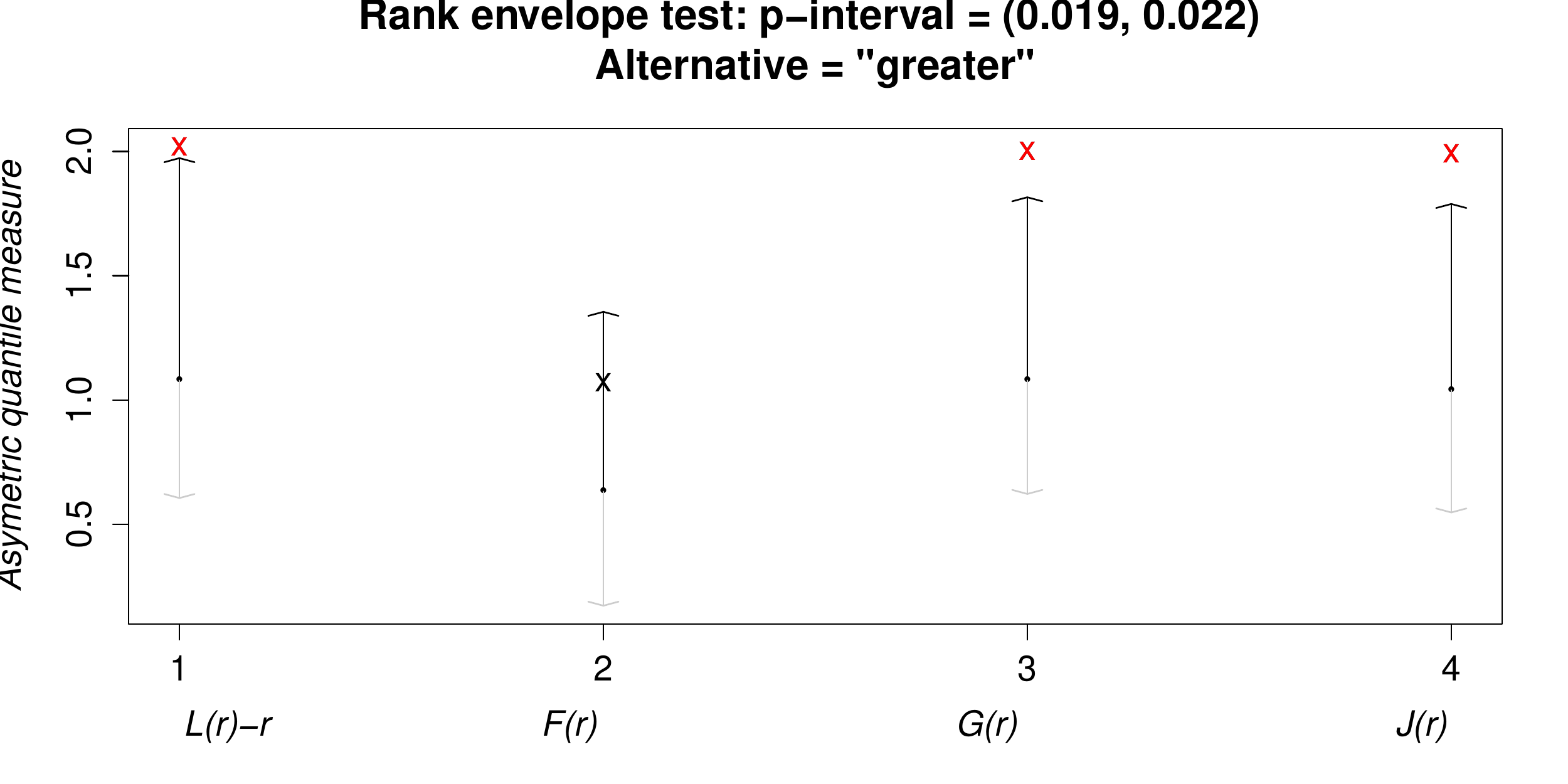} }
\captionof{figure}{\label{testLFGJAsQ} The rank test used for the combination of four scaled deviation measures of $L(r)-r$, $F(r)$, $G(r)$ and $J(r)$ functions performed with $s=999$ simulations of the null model.}
\end{figure}
\end{center}

Further, Figure \ref{testLFGJAsQ} shows the rank test used for the combination of four scaled \citep[asymmetric quantile, see][]{MyllymakiEtal2015a} maximum deviation measures of  $L(r)-r$-, $F(r)$-, $G(r)$- and $J(r)$-functions for testing the same situation as in Figure \ref{testLFGJ}.
We can see that the rejection of this test is due to $L$-, $G$- and $J$-functions, but based on this test we do not know the reason of rejection.

We investigated by a simulation study how the power of the rank envelope test and the scaled maximum deviation test \citep{MyllymakiEtal2015a} is affected by using one or more test functions.
Since the results for the deviation test were similar to those of the rank envelope test, in the following we show and discuss only the latter.

\subsection{Simulation study}

In our previous study \citet{MyllymakiEtal2015b}, we studied the power of rank envelope and deviation tests with $L$- and $J$-functions. We found out, as expected, that which of the test functions is more powerful depends on the null and true models.
Now we extend the previous study to the tests with several tests functions. We would like to show, that the empirical type I error probability stays at the desired level in the combined test and that the power of the combined test is not much smaller than the power of the test with the most powerful test function.

In addition to $L$- and $J$-functions, we add to the study the empty space function ($F$-function) and the nearest neighbour distribution function ($G$-function). As tests functions we use standard estimators of these summary functions \citep[see e.g.][]{IllianEtal2008}.

\subsubsection{Design of the study}

 We used the following point process models
\begin{enumerate}
 \item Poisson process with intensity $\lambda$, i.e.\ complete spatial randomness (CSR),
 \item Strauss($\beta$, $\gamma$, $R$) process, where $R>0$ is the interaction radius, and $\beta>0$ and $0\leq \gamma\leq 1$ control the intensity and the strength of interaction, respectively,
 \item {\em Mat{\'e}rn cluster process} MatClust($\lambda_p$, $R_d$, $\mu_d$), where
 $\lambda_p$ is the intensity of parent points, and $R_d$ and $\mu_d$ specify the cluster radius and mean number per cluster for the daughter points,
 \item {\em non-overlapping Mat{\'e}rn cluster process} NoOMat\-Clust($\lambda_p$, $R_d$, $\mu_d$, $R$),
where the parent points follow a hard-core process with hard-core distance $R$ (i.e.\ Strauss($\beta$, 0, $R$) process), and
 \item {\em mixed Mat{\'e}rn cluster process} (MixMatClust), which is a superposition of two Matern cluster processes.
\end{enumerate}
For further details on the processes see e.g.\ \citet{IllianEtal2008}. The first three processes were used as null models. The employed null and true models are specified more precisely below in Tables \ref{Table:signlevels1} and \ref{Table:powers}.

The chosen true model was simulated $1000$ times in the unit square. Then the parameters of the null model were estimated for each simulation. For the CSR hypothesis, only the intensity of the tested point pattern was estimated. The parameters of the Mat{\'e}rn cluster process were estimated by the minimum contrast estimation based on the pair correlation function (the non-cumulative counterpart of the $L$-function), whereas for the Strauss process the the logistic likelihood method \citep{BaddeleyEtal2014} was used.

Then $s=1999$ simulations of the fitted null model were done and the rank test with extreme rank count ordering was applied to each combination of test functions $(L, F, G, J)$.   
For the sake of compactness, the tables in the following show only results for the combinations where the most commonly used function, the $L$-function, is involved.
For each combination of test functions and each model, we calculated the proportion of rejections
of the null model among the $1000$ simulations (rejection if $p\leq 0.05$).

\subsubsection{Empirical type I error probabilities of the combined rank test for a simple null model}

First, we studied the type I error probabilities for the CSR, Strauss($350$, $0.4$, $0.03$) and MatClust($50$, $0.06$, $4$) models. The latter two models deviate stronly from CSR and they are similar to the null models used in the power study.
The parameters of the null model were assumed to be known. All the estimated type I error probabilities are close to the nominal level $\alpha=0.05$, see Table \ref{Table:signlevels1}.
Indeed, for $\alpha=0.05$, the proportions of rejections should be in
the interval $(0.037, 0.064)$ with the probability 0.95 (given by the 2.5\% and
97.5\% quantiles of the binomial distribution with parameters 1000 and 0.05).
Thus, we conclude that the rank test has correct empirical type I error probability for any combination of test functions.

{\footnotesize
\begin{table}
\captionof{table}{Estimated type I error probabilities of the rank test with different combinations of test functions. The parameters of the null model are known.}\label{Table:signlevels1}
\centering
\fbox{%
\begin{tabular}{l|ccccccccc}
Simulated model &  $L$  & $F$  & $G$  & $J$  & $L,F$ &  $L,G$  & $L,J$  & $L,G,J$  & $L,F,G,J$ \\\toprule
Poisson(200) &  0.047 &	0.046	&0.049	&0.044	&0.049	&0.046	&0.044	&0.043	&0.043  \\
Strauss(350, 0.4, 0.03) &0.047	&0.056&	0.045&	0.052&	0.051&	0.045&	0.052& 0.052
&0.056 \\
MatClust(50, 0.06, 4) & 0.064	&0.047&	0.057&	0.048&	0.056&	0.053&	0.050&	0.050&	0.051\\
\end{tabular}}
\end{table}
}
\subsubsection{Effect of overfitting - type I error probabilities of the combined rank test for a composite null model}

Practically, the true parameters are unknown and have to be estimated. In such a case, the estimated type I errors are often appropriate for such test functions which are only loosely correlated with the fitting procedure \citep{Diggle2013, MyllymakiEtal2015b}. Based on our study, see Table \ref{Table:signlevels2}, they are the $L$- and $J$-function for the test of complete spatial randomness and $J$ for the Strauss process. For the Mat\'ern cluster process, the $J$-function is also less conservative than the other functions. The test function $F$ seems to be very conservative in all cases. 
Table \ref{Table:signlevels2} further shows that combining functions that are only loosely correlated with the estimation procedure and and the ones that are highly correlated averages the level of conservativeness of the test.

{\footnotesize
\begin{table}
\captionof{table}{Estimated type I error probabilities of the rank test with different combinations of test functions. The parameters of the null model are fitted.}\label{Table:signlevels2}
\centering
\fbox{%
\begin{tabular}{l|ccccccccc}
Simulated model &  $L$  & $F$  & $G$  & $J$  & $L,F$ &  $L,G$  & $L,J$  & $L,G,J$  & $L,F,G,J$ \\\toprule
Poisson(200) &  0.054 &  0.000 &  0.022 &  0.041 &  0.045 &  0.036  & 0.042 & 0.035 & 0.033  \\
Strauss(350, 0.4, 0.03) & 0.025	&0&	0.015&	0.015&	0.019&	0.022&	0.021& 0.017
& 0.017 \\
MatClust(50, 0.06, 4) & 0.008  & 0.000  & 0.014  & 0.032  & 0.005  & 0.012  & 0.022  & 0.019  & 0.021\\

\end{tabular}}
\end{table}
}

\subsubsection{Rejection rates of the combined rank test}

\begin{center}
\begin{figure}[htbp]
\includegraphics[scale=0.3]{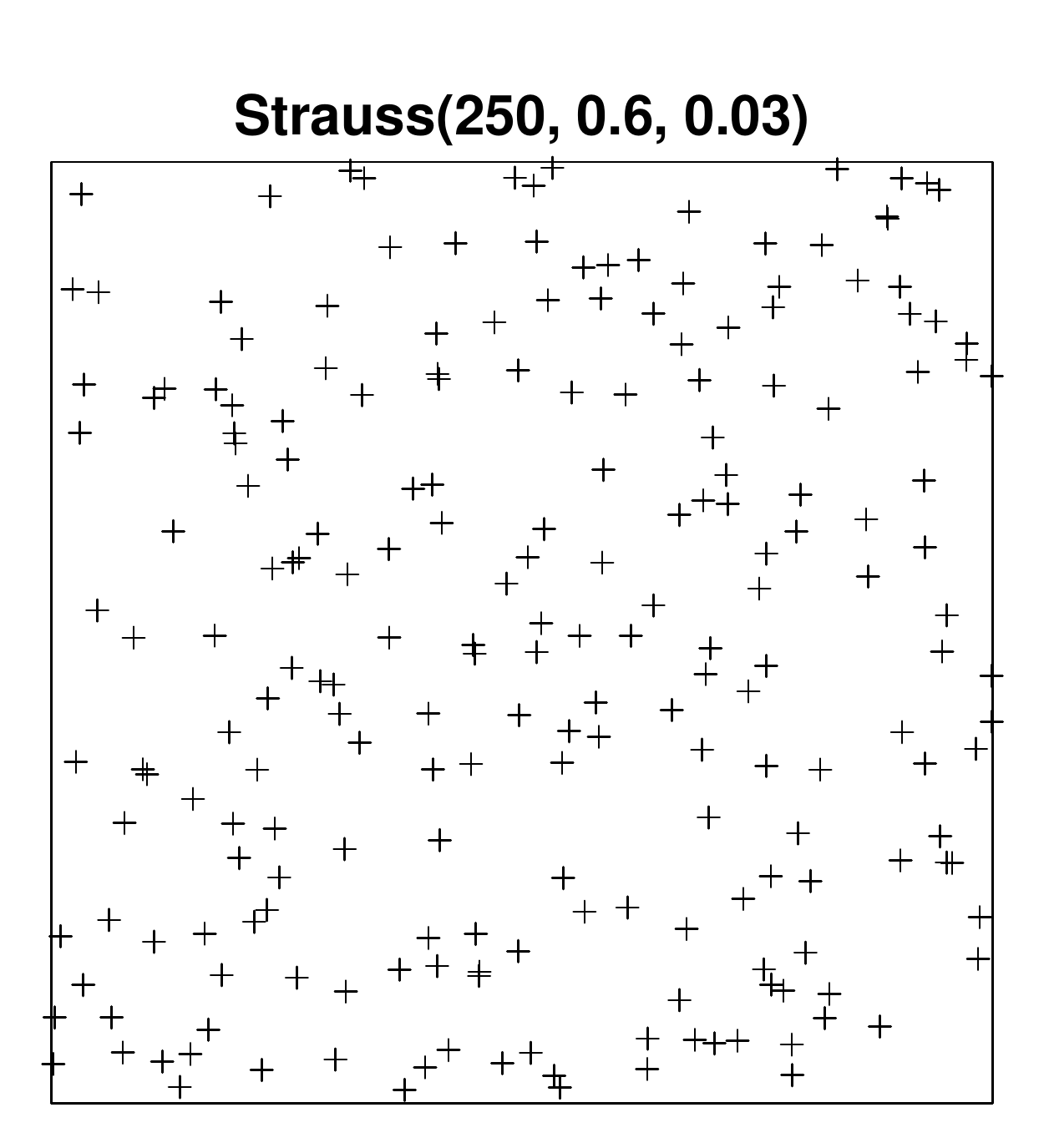} \hspace{1cm} \includegraphics[scale=0.3]{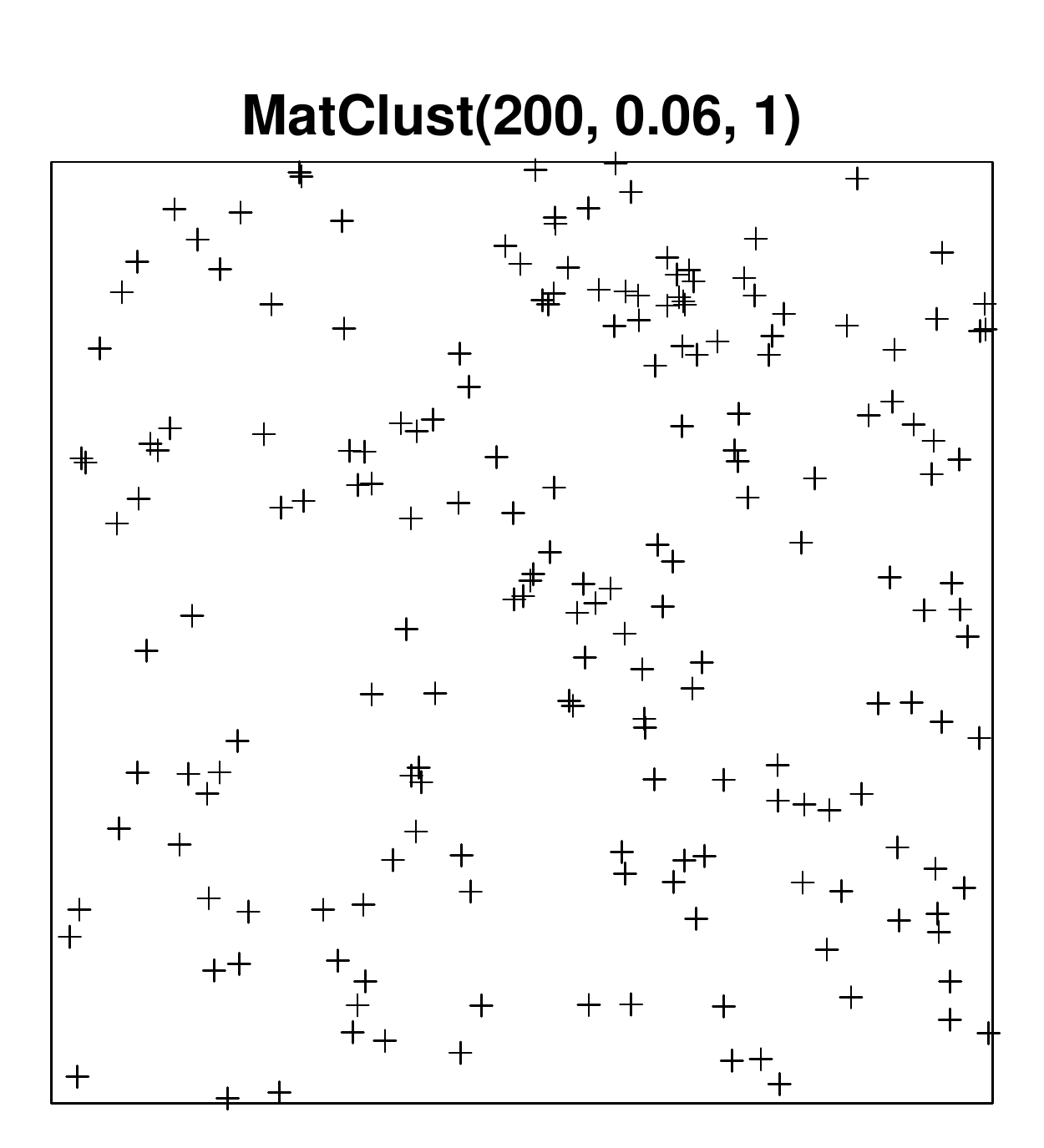} \hspace{1cm} \includegraphics[scale=0.3]{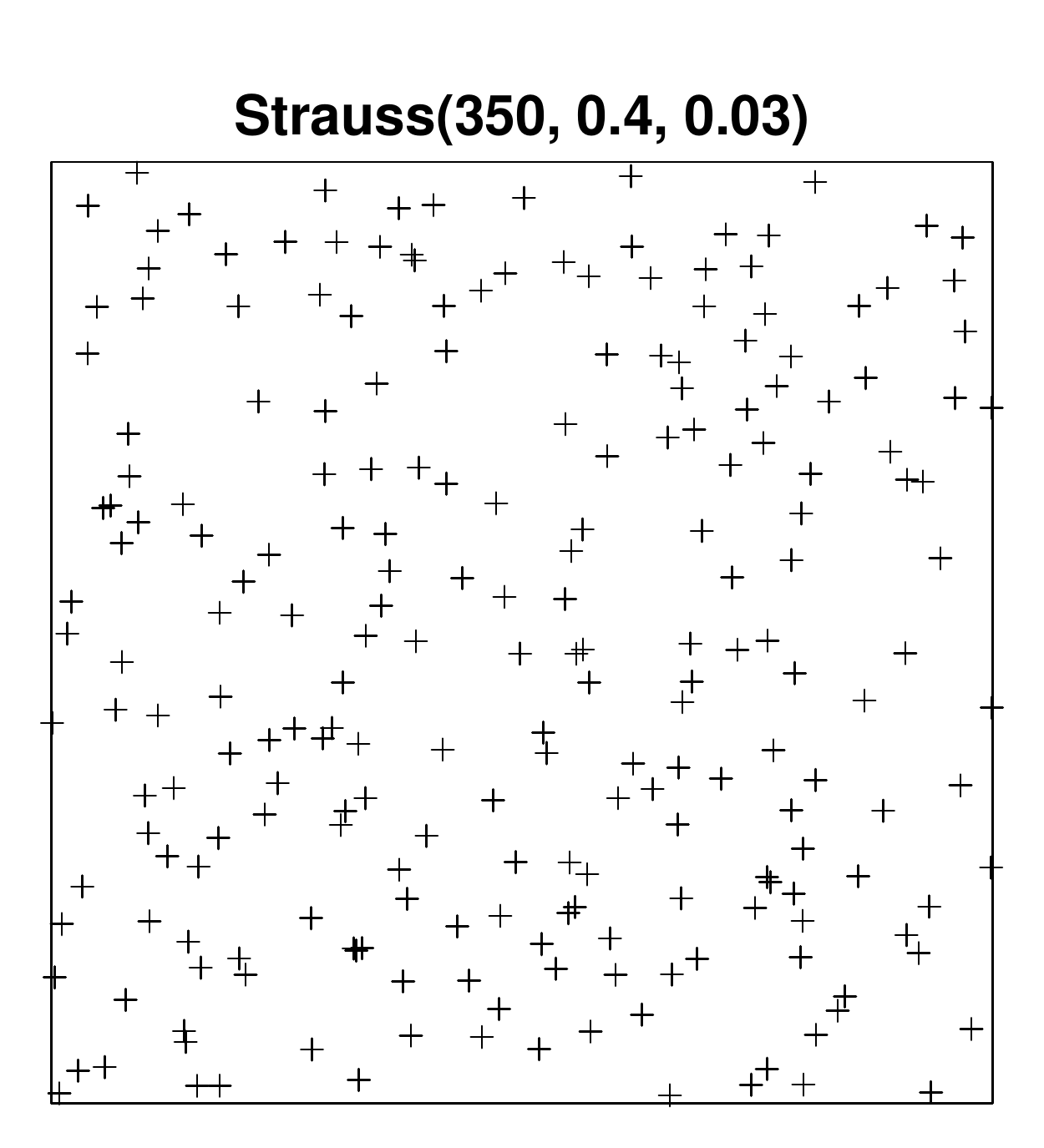} \\
\includegraphics[scale=0.3]{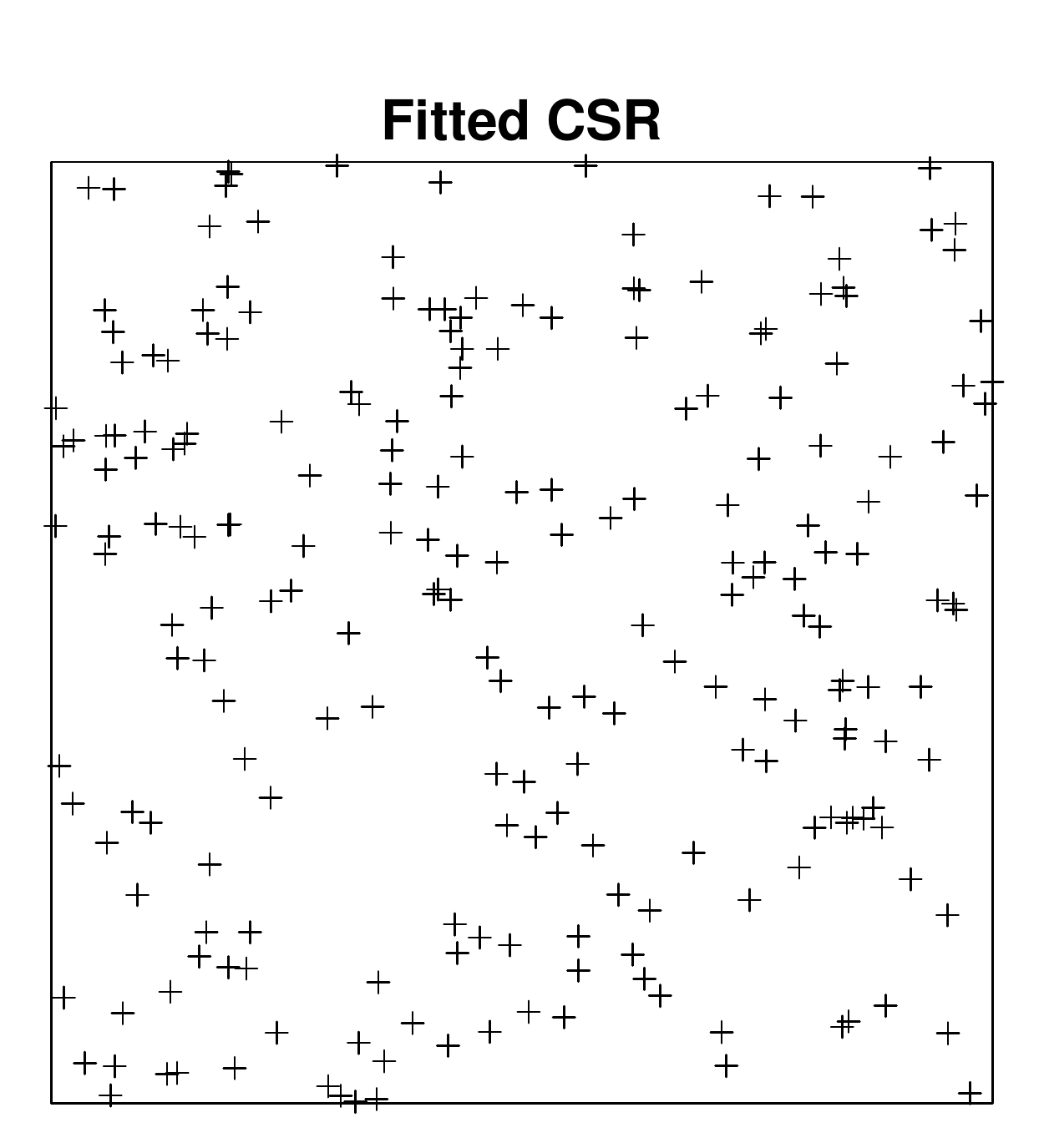} \hspace{1cm} \includegraphics[scale=0.3]{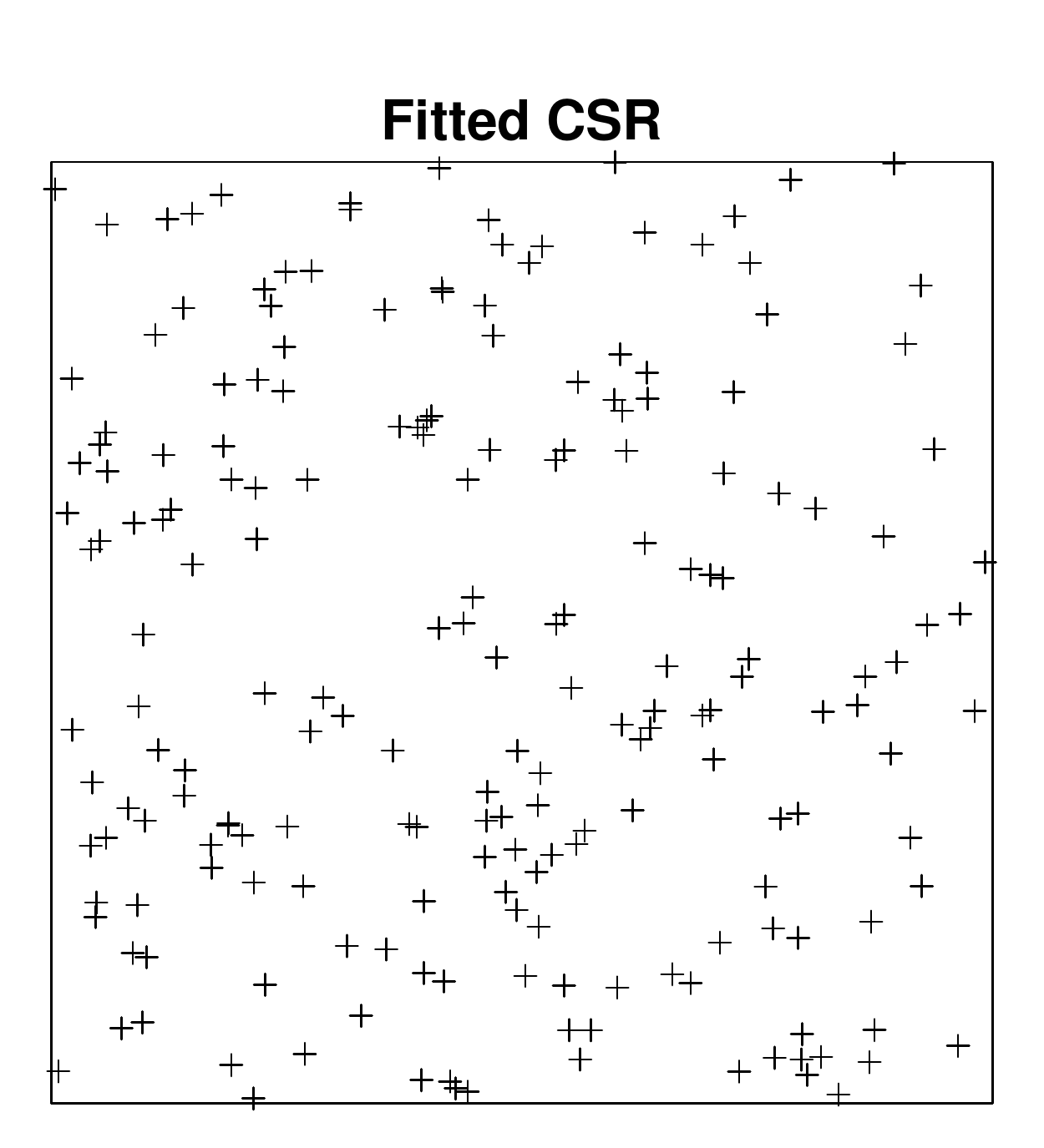} \hspace{1cm} \includegraphics[scale=0.3]{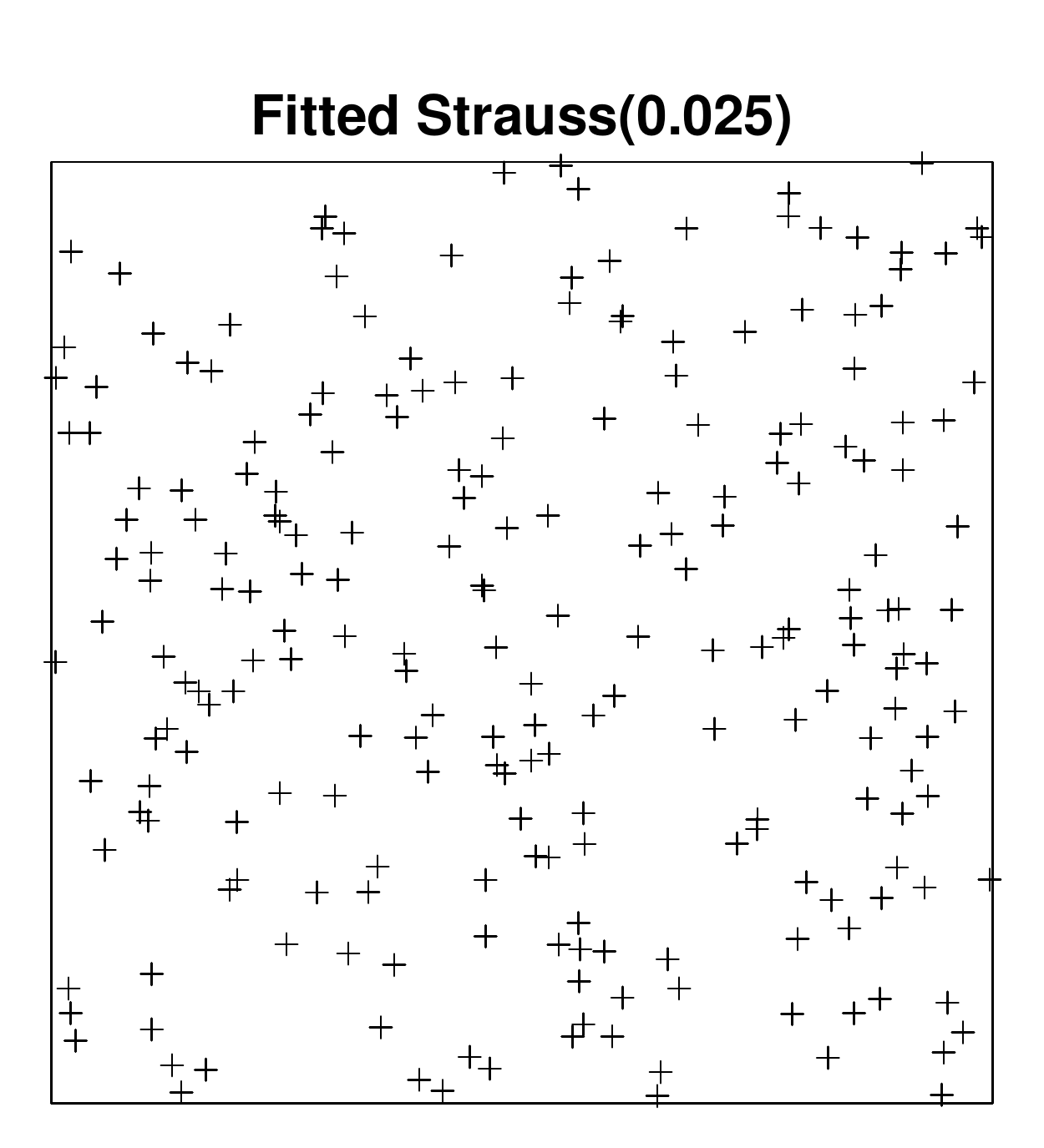} \\
\includegraphics[scale=0.3]{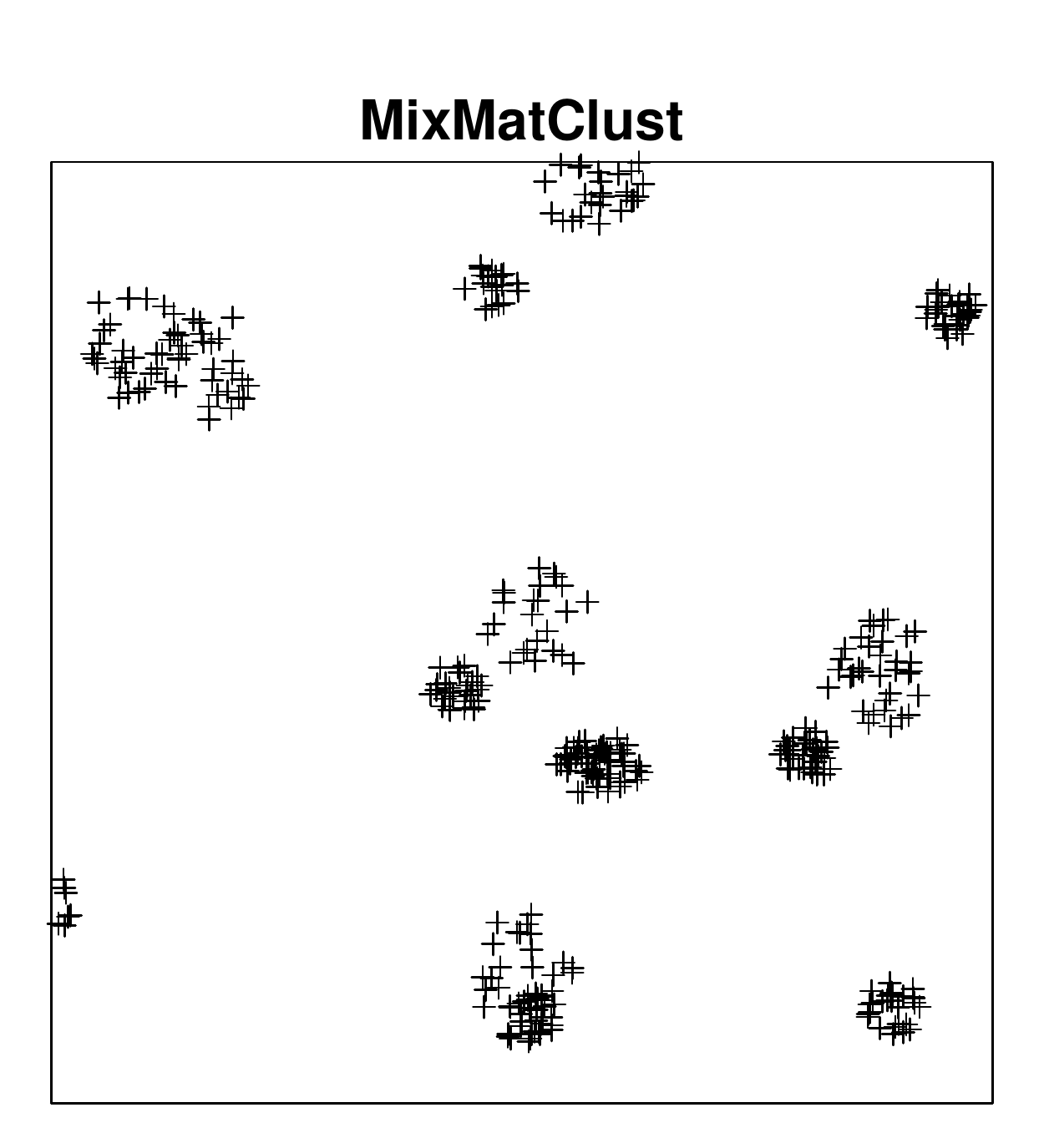} \hspace{1cm} \includegraphics[scale=0.3]{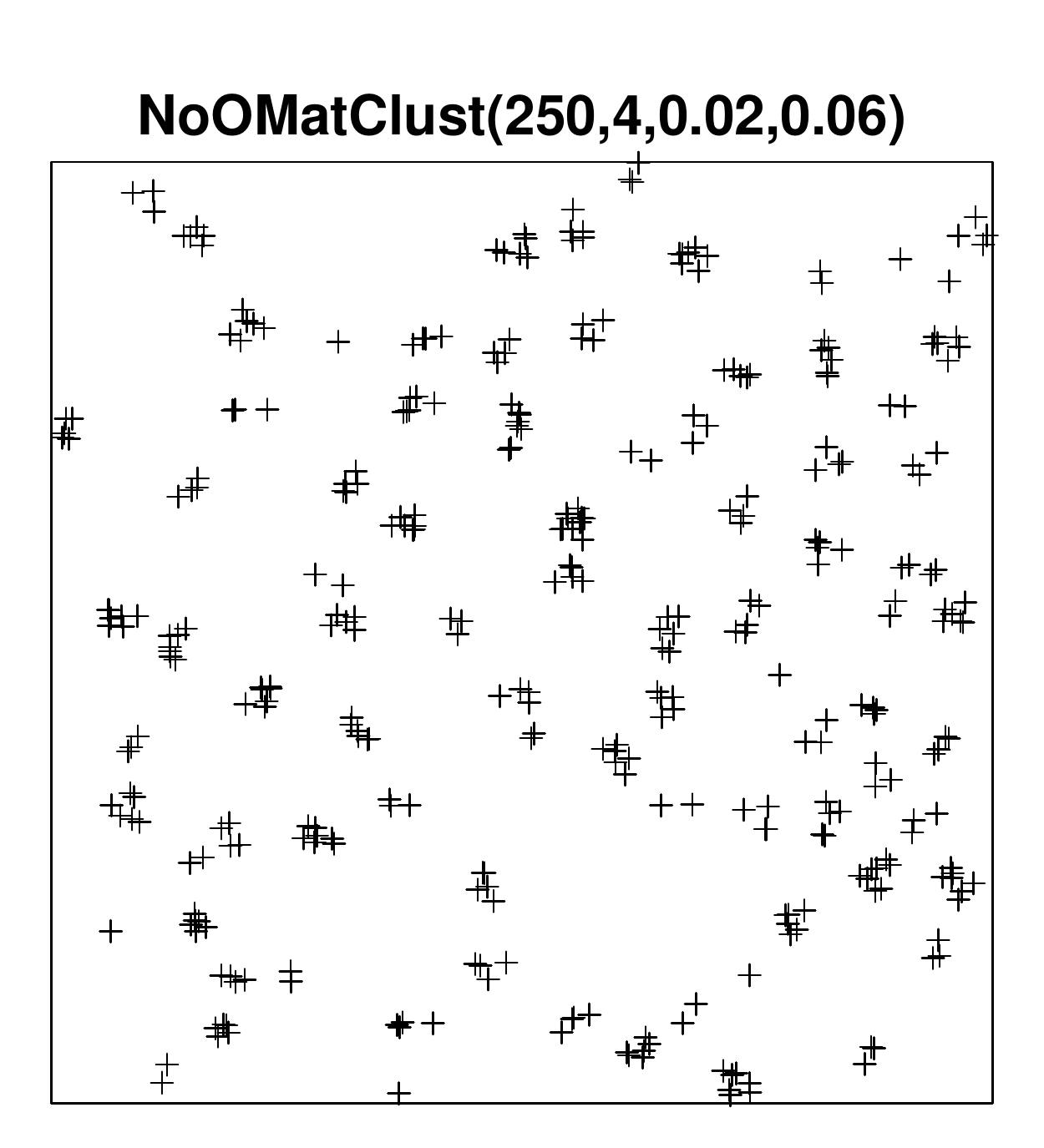} \\
\includegraphics[scale=0.3]{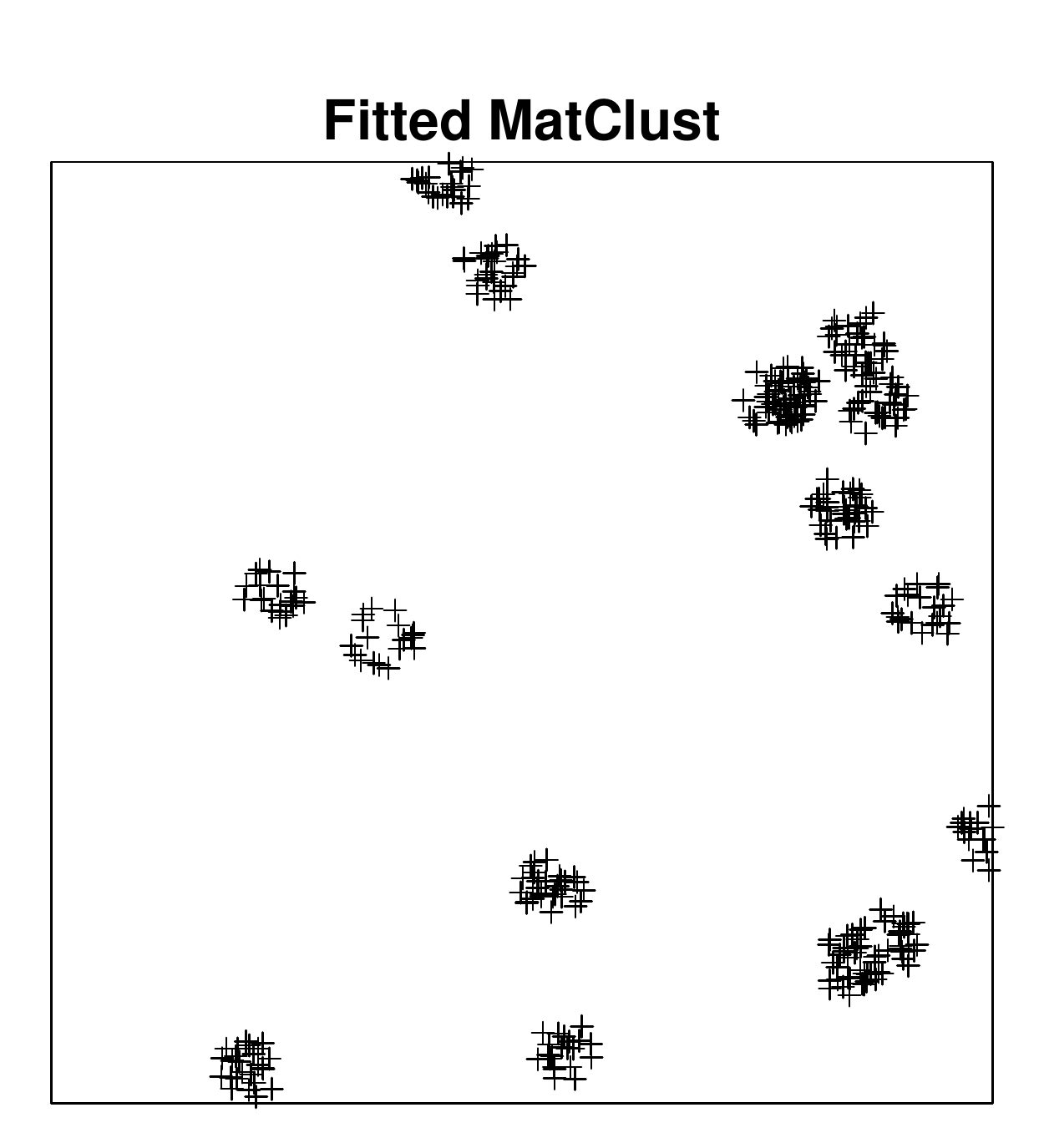} \hspace{1cm} \includegraphics[scale=0.3]{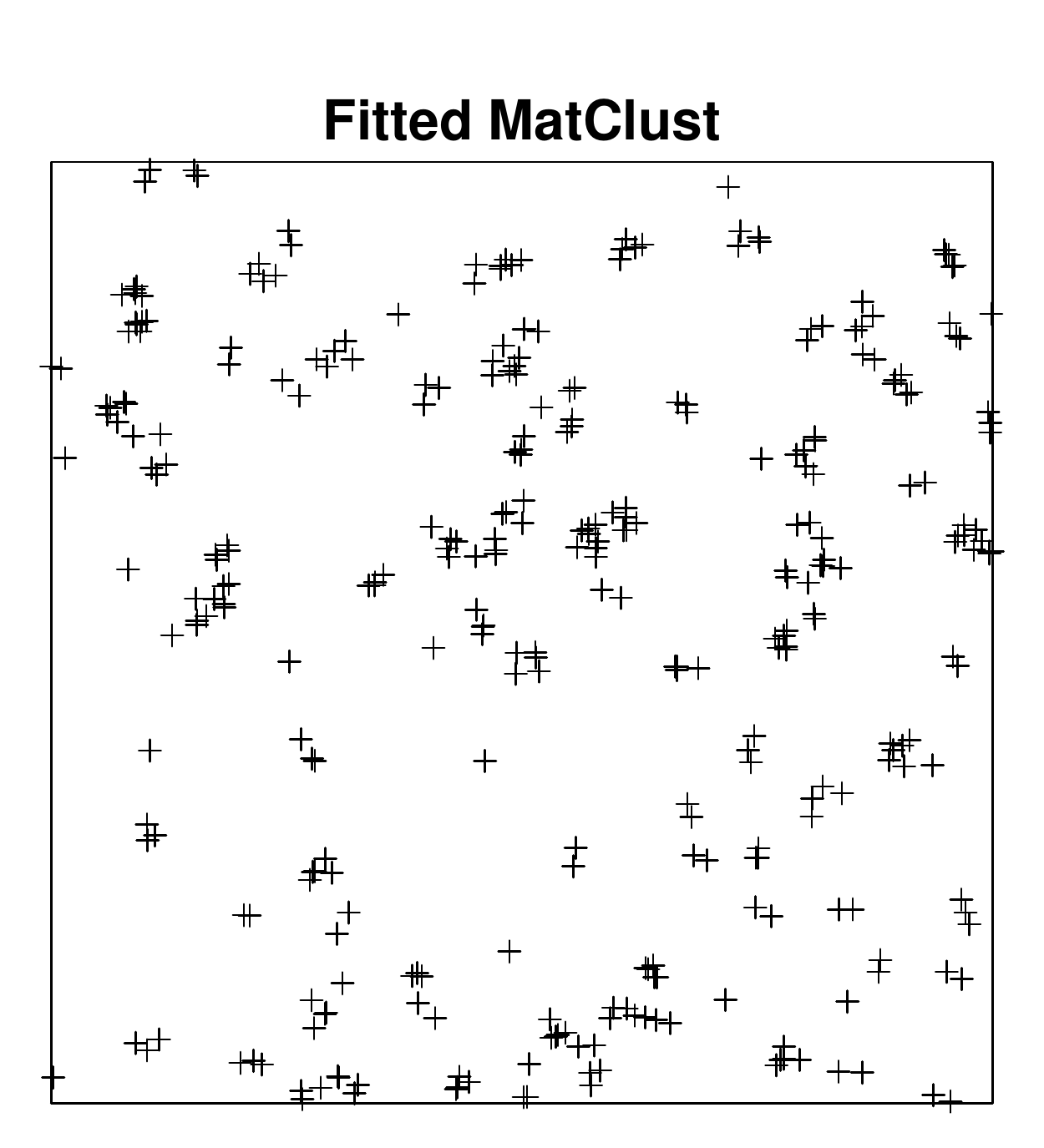}
\captionof{figure}{\label{realizations} First and third lines show realizations of chosen true models, while second and fourth lines show its fitted null models. Mixed matern cluster model is superposition of MatClust(10, 0.06, 30) and MatClust(10, 0.03, 30).}
\end{figure}
\end{center}

The null and true models are shown in the two leftmost columns of Table \ref{Table:powers}.
The parameters of the alternative models were chosen such that the deviation from the null model is moderate.
One realization of each studied alternative model with its fitted null model are shown in Figure \ref{realizations} for the illustration of closeness of null and true models.
Table \ref{Table:powers} also shows the obtained rejection rates of the rank test with various combinations of test functions. We observed the following:
\begin{enumerate}
 \item The combined test has only a bit lower power than the most powerful test statistic in all studied cases.
 \item Different single test functions can lead to very different powers (see e.g.\ line 5 of Table \ref{Table:powers}).
 \item The last line of Table \ref{Table:powers} shows that 
 using a highly conservative test function (as $L$ for the Mat{\'e}rn cluster process) together with a less conservative test function (here $J$) can increase the power with respect to using only the less conservative test function.
 \item Finally, the last column of Table \ref{Table:powers} shows that even by adding $F$-function, which has just very week power, the power decreases only little.
\end{enumerate}

{\scriptsize
\begin{table}
\captionof{table}{Estimated powers of the rank test with different combinations of test functions. The Strauss null model was fitted with $R=0.02$.}\label{Table:powers}
\centering
\fbox{%
\begin{tabular}{l|l|ccccccccc}
True model &  Null & $L$  & $F$  & $G$  & $J$  & $L,F$ &  $L,G$  & $L,J$  & $L,G,J$  & $L,F,G,J$ \\\toprule
Strauss(250, 0.6, 0.03) & CSR &  0.622 &  0.010 &  0.428 &  0.615 &  0.591 &  0.553 &  0.608 &  0.566 &  0.553 \\
MatClust(200, 0.06, 1)  & CSR &  0.789 &  0.208 &  0.377  & 0.573 &  0.772 &  0.744 &  0.773 &  0.747 &  0.737 \\
Strauss(350, 0.4, 0.03) & Strauss &0.816	&0.012	&0.585	&0.721	&0.781	&0.746&	0.768& 0.712&0.691\\
MixMatClust             & MatClust &  0.000 &  0.000 &  0.949 & 0.949 &  0.000  & 0.944 &  0.944  & 0.944  & 0.944 \\
NoOMatClust             & MatClust &  0.537 &  0.000  & 0.185  & 0.267  & 0.490  & 0.448  & 0.488  & 0.439  & 0.424 \\
\end{tabular}}
\end{table}
}

This simulation study tells us, that we can generally construct a rank test which is sensitive to ``all'' possible alternative hypotheses by joining several test functions without worry of loosing the power of the test.

\section{Simultaneous goodness-of-fit test for several point patterns}\label{sec:gof-several-patterns}

%

\begin{figure}[h!]
\centering
\includegraphics[width=0.25\linewidth]{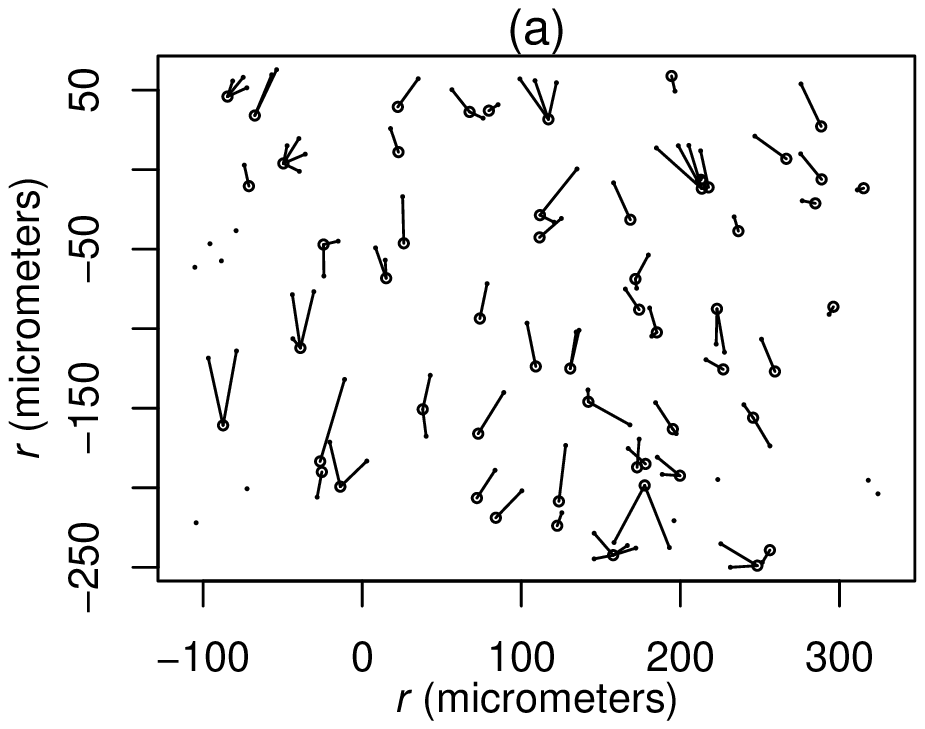}\includegraphics[width=0.25\linewidth]{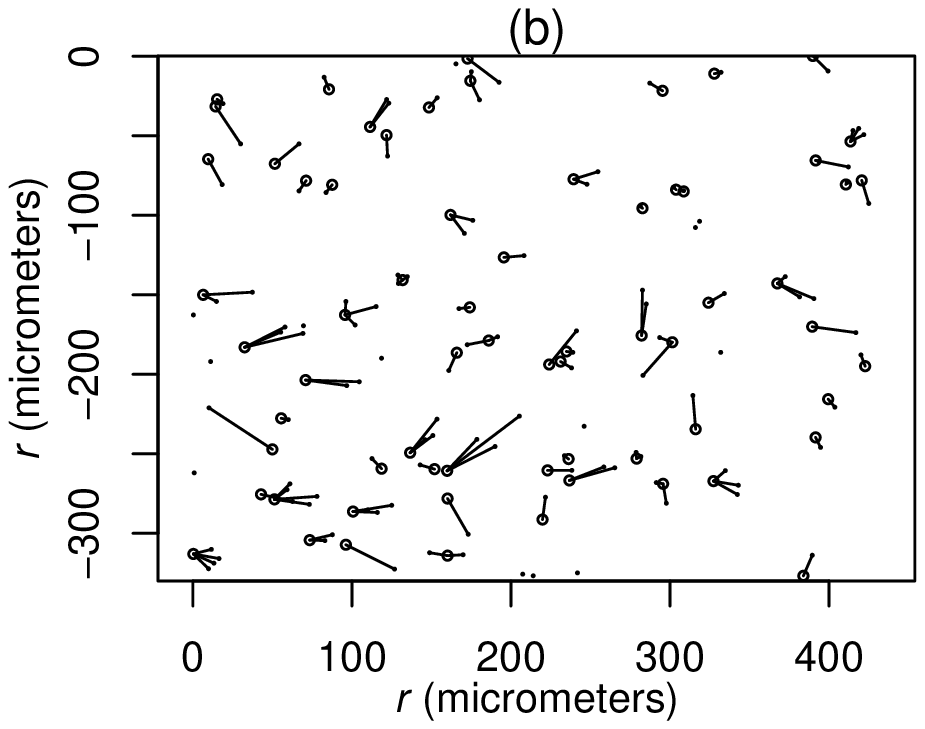}\includegraphics[width=0.25\linewidth]{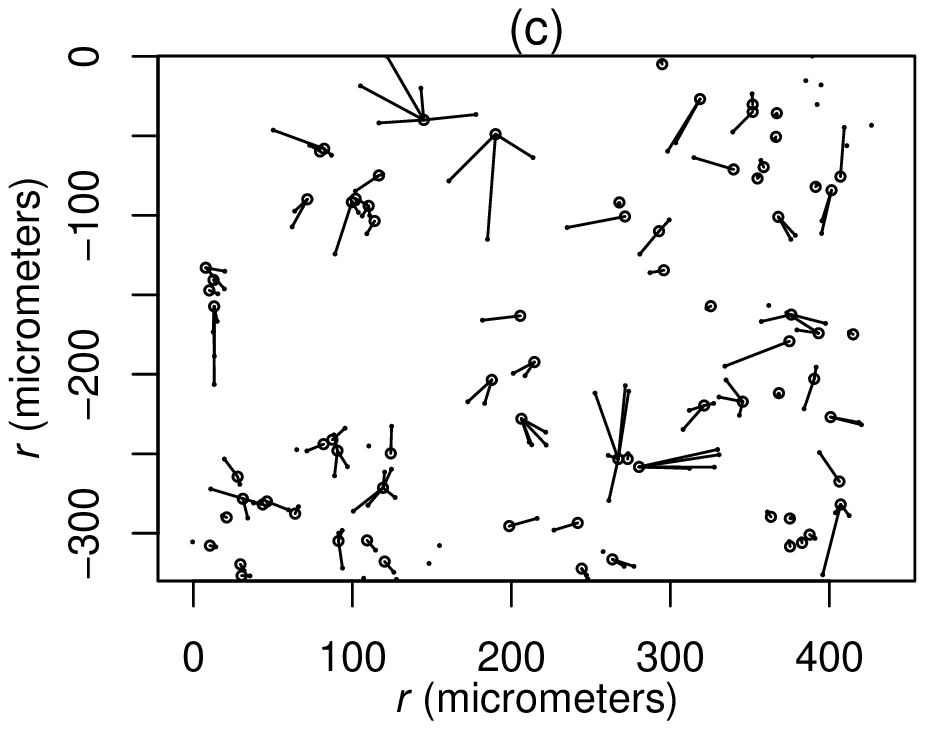}\includegraphics[width=0.25\linewidth]{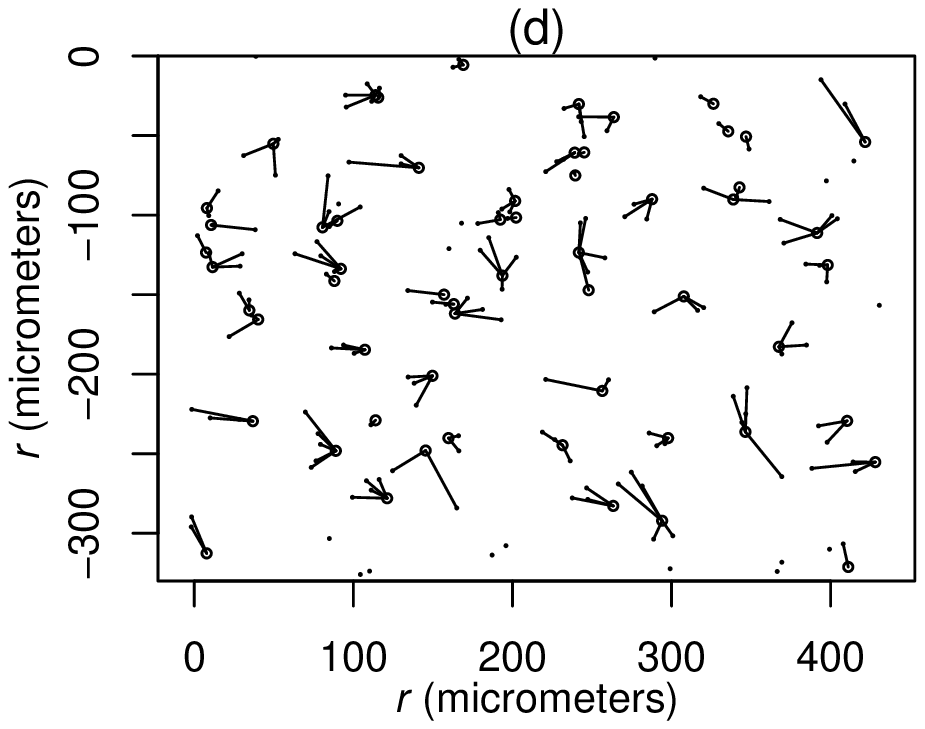}
\captionof{figure}{\label{Fig_ENFs}Epidermal nerve fiber patterns, where fibers are replaced by line segments connecting the end points (small black dots) and the entry points (black circles). Subjects: (a) 171, (b) 224, (c) 230 and (d) 256.}
\end{figure}

Figure \ref{Fig_ENFs} shows point patterns of entry and end points of
epidermal nerve fibers (ENFs) that were previously analysed in \citet{OlsboEtal2013}.
The ENFs are thin nerve fibers living in the outmost layer of the skin called epidermis.
While \citet{KennedyEtal1996} reported diminished numbers of ENFs in subjects suffering
from diabetic neuropathy, \citet{WallerEtal2011} and \citet{MyllymakiEtal2014} tried
to quantify increased clustering of ENFs in such subjects based on spatial second-order analysis.
\citet{OlsboEtal2013} further proposed preliminary point process models for the entry and
end points based on data from thigh of healthy subjects.

We tested the CSR hypothesis for the entry and end point patterns in Figure \ref{Fig_ENFs},
as was done in \citet{OlsboEtal2013} as the first step in analysing the data sets.
While in \citet{OlsboEtal2013} the CSR hypothesis was tested separately for each pattern
by means of the refined envelope test proposed by \citet{GrabarnikEtal2011},
we now performed the test jointly for all the entry point patterns and for all the end point patterns.
As \citet{OlsboEtal2013}, we used as the test function an estimator of the $L$-function
with translational edge correction \citep[see e.g.][]{IllianEtal2008}.
We performed a two-sided rank envelope test on the interval of distances $I=[0, 80]$ (micrometers).

\begin{figure}[h!]
\centering
\includegraphics[width=1\linewidth]{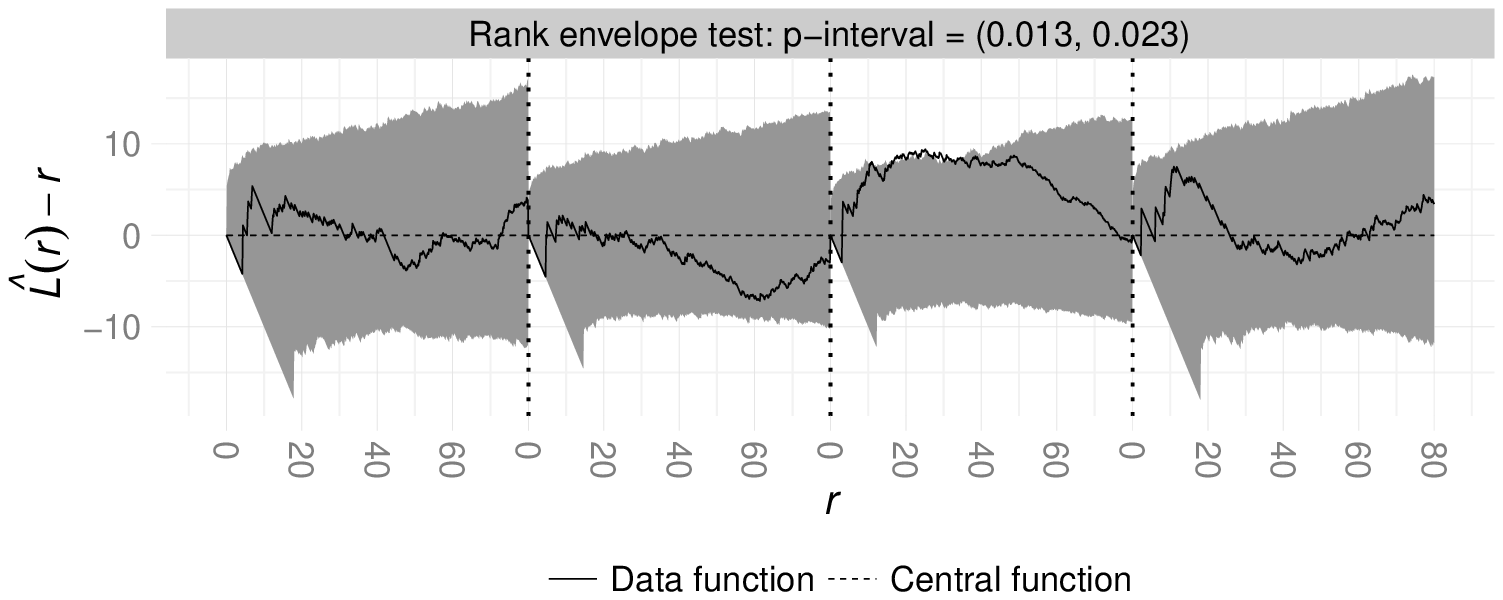}
\captionof{figure}{\label{Fig_ENFs_rank_base}Rank envelope test for testing CSR of the ENF entry point patterns in Figure \ref{Fig_ENFs}. The number of simulations is $s=20000$ and $T(r) = \hat{L}(r)$.}
\end{figure}

\begin{figure}[h!]
\centering
\includegraphics[width=1\linewidth]{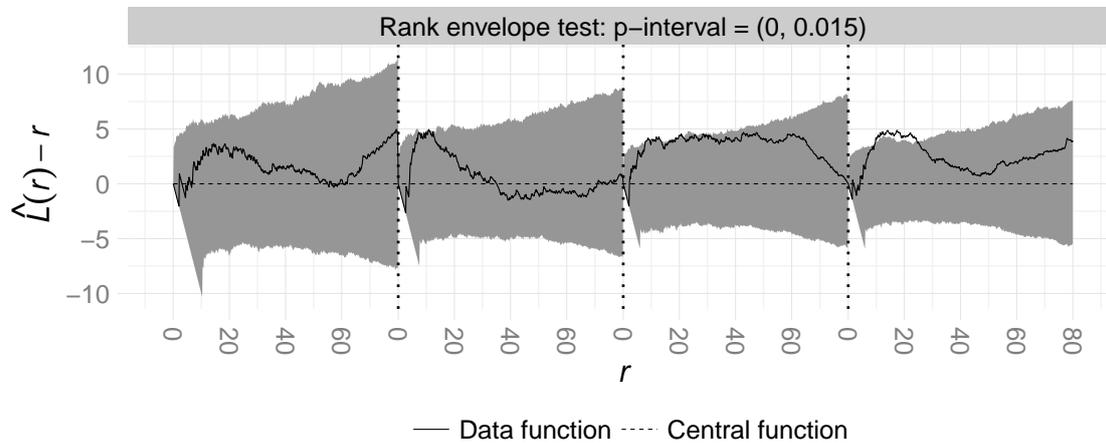}
\captionof{figure}{\label{Fig_ENFs_rank_end}Rank envelope test for testing CSR of the ENF end point patterns in Figure \ref{Fig_ENFs}. The number of simulations is $s=20000$ and $T(r) = \hat{L}(r)$.}
\end{figure}

The combined rank envelope test ($s=20000$) rejects the CSR hypothesis both for the entry and end point patterns, see Figures \ref{Fig_ENFs_rank_base} and \ref{Fig_ENFs_rank_end}.
The reason of rejection for the entry points is the pattern of Subject 230.
For end points, the rejection is due to the three subjects 224, 230 and 256.
The same was in fact concluded in \citet{OlsboEtal2013}.
However, here we do only one test with a global type I error probability instead of four tests and provide a $p$-value for this combined test. 
The $p$-values based on extreme rank count ordering are 0.016 and 0.0036 for the entry and end points, respectively.

The number of performed simulations $s=20000$ is obviously large.
We could have performed the test based on the extreme rank count ordering using a smaller number of simulations in order to have smaller computational load.
For the entry points, the extreme rank count $p$-value obtained with $s=4999$ is $0.0014$,
while the $p$-interval is $(<0.001, 0.040)$.
For the end points, the corresponding $p$-value and -interval are $0.0022$ and $(<0.001, 0.047)$.
So, we in fact come to the same conclusions with $s=4999$ as with $s=20000$ in this case (figures not shown).

\section{Comparison of groups of point patterns (ANOVA)}\label{sec:ANOVA}

In this section, we describe how the rank test can be used to compare groups of point patterns via a chosen test function which is computed for every point pattern. This task leads to a functional one way ANOVA problem, which was already solved by many authors. For example \citet{CuevasEtal2004} introduced asymptotic version of the ANOVA $F$-test, \citet{RamsaySilverman2006} describe a bootstrap procedure based on pointwise $F$-tests, \citet{AbramovichAngelini2006} used wavelet smoothing techniques, \citet{FerratyEtal2007} used dimension reduction approach and \citet{CuestaFebrero2010} used several random univariate projections on which the $F$-test is applied and then the tests are bounded together through false discovery rate. The last procedure was applied to a point pattern data of colorectal tumors in \citet{AliyEtal2013}. There is also a possibility to transform the function into one number and apply a classical anova but such procedures can be blind against some alternatives. 

In the point patterns literature the group comparison is done either using functional anova as in \citet{AliyEtal2013} or using a bootstrap procedure as in \citet{DiggleEtal1991},  \citet{DiggleEtal2000} or \citet{SchladitzEtal2003}. Furthermore, \citet{Hahn2012} proposed a pure permutation procedure to correct the inaccurate significance level of the bootstrap procedure. In these works the univariate statistic summarizing the overall differences between the groups is used and permuted or bootstrapped. 

In Section \ref{sec:fANOVA} we describe how the rank test can be used to solve the general ANOVA problem. 
An advantage of our proposed rank method is the resulted graphical interpretation of the results: it directly identifies the distances which are responsible for the potential rejection. Another advantage is that the rank test and, thus, the proposed functional ANOVA test is performed exactly with the desired significance level. 

In Section \ref{sec:one-way-group-comparison} we investigate the possibility to use the rank test for determining the differences between groups of functions. The rank test can be used to determine which group differences and which distances are responsible for the possible rejection.

\begin{center}
\begin{figure}[htbp]
\centerline{\includegraphics[scale=0.30]{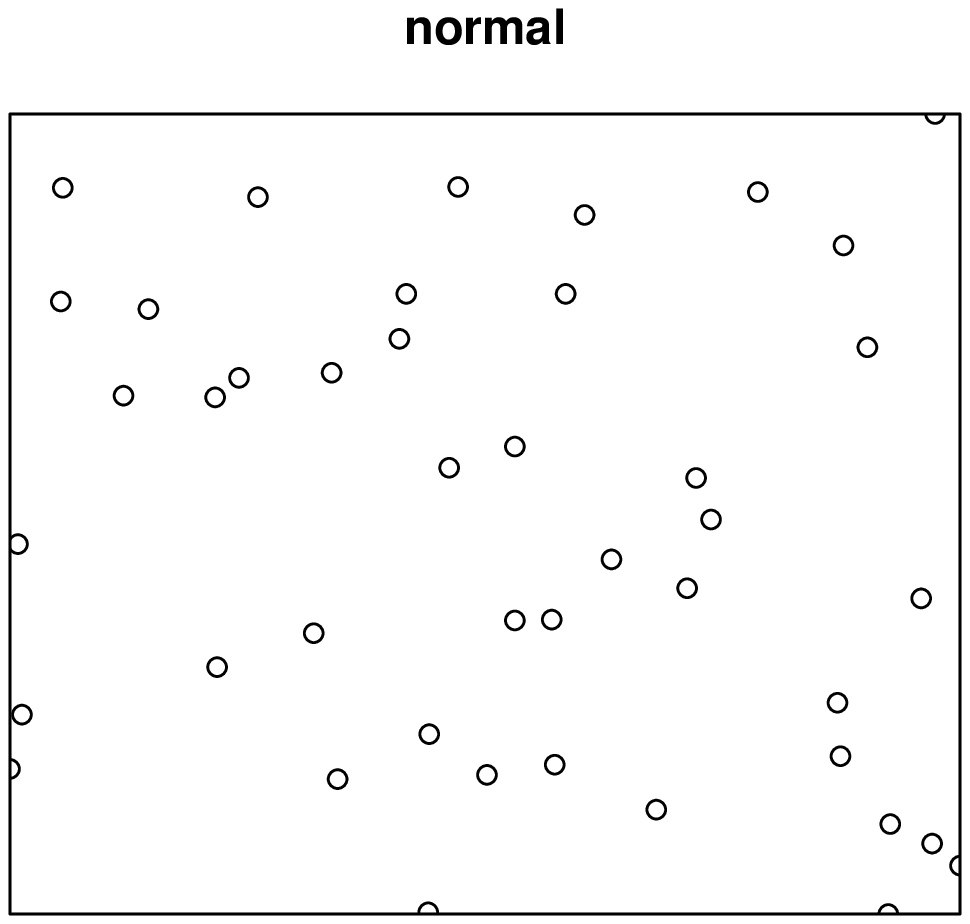} \includegraphics[scale=0.30]{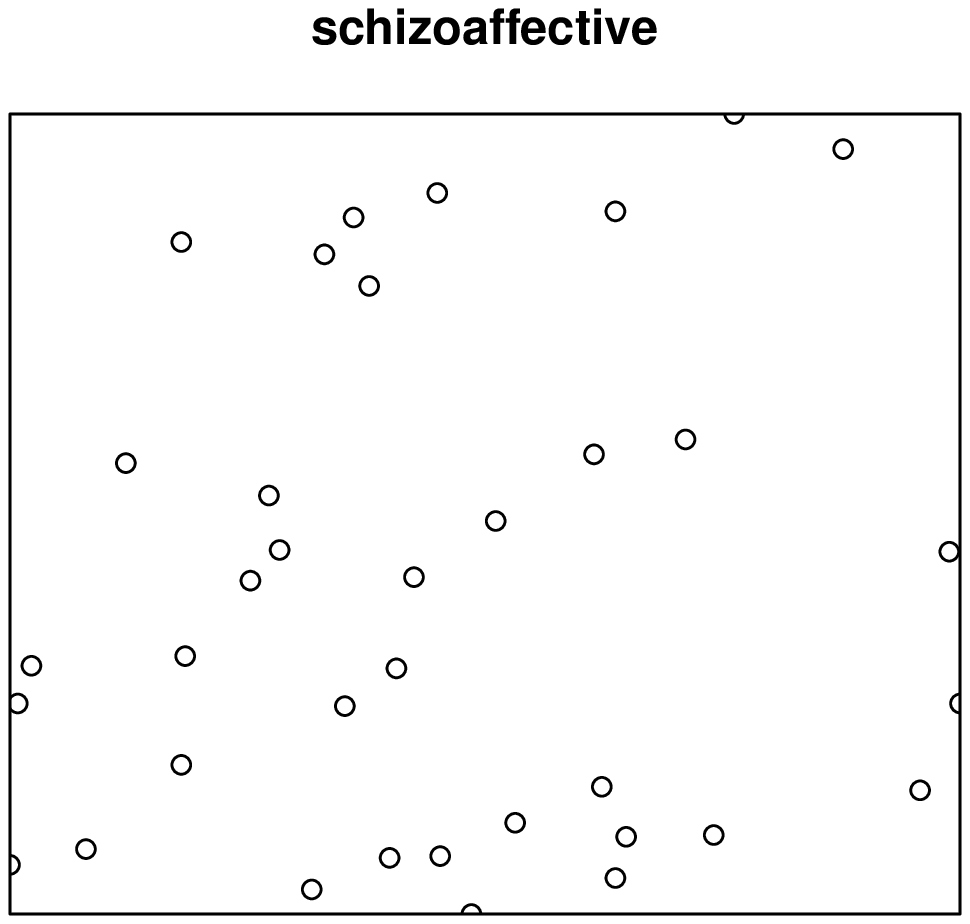} \includegraphics[scale=0.30]{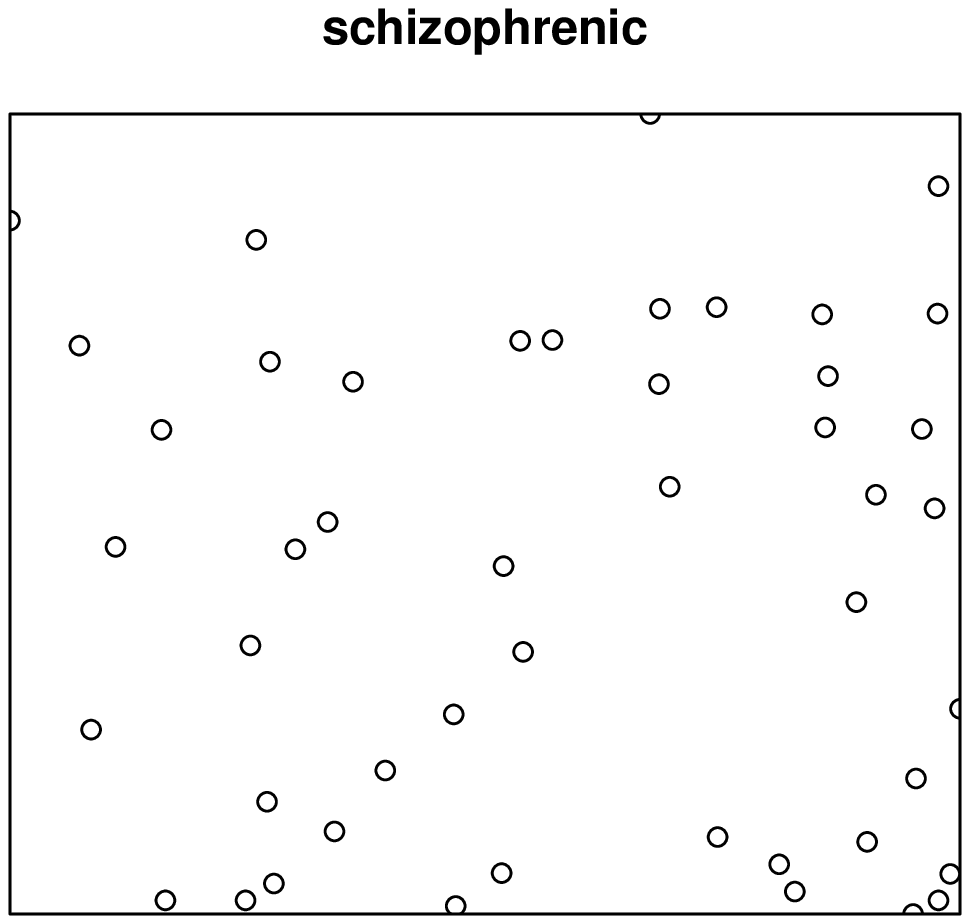}}
\captionof{figure}{\label{fig:neurons} One representative point pattern of each group of pyramidal neuron positions.}
\end{figure}
\end{center}

To describe our approach we reanalyse the data of  \citet{DiggleEtal1991} containing 3 groups of pyramidal neurons in the cingulate cortex of humans, the normal - control group, schizoaffective group and schizophrenic group. One representative pattern from each group can be seen in Figure \ref{fig:neurons}.  (We discarded the point patterns with less than 20 points prior to the analysis, which led to 12 point patterns in the normal group, 7 patterns in the schizoaffective group and 7 patterns in the schizophrenic group). 

\begin{center}
\begin{figure}[htbp]
\centerline{\includegraphics[scale=0.45]{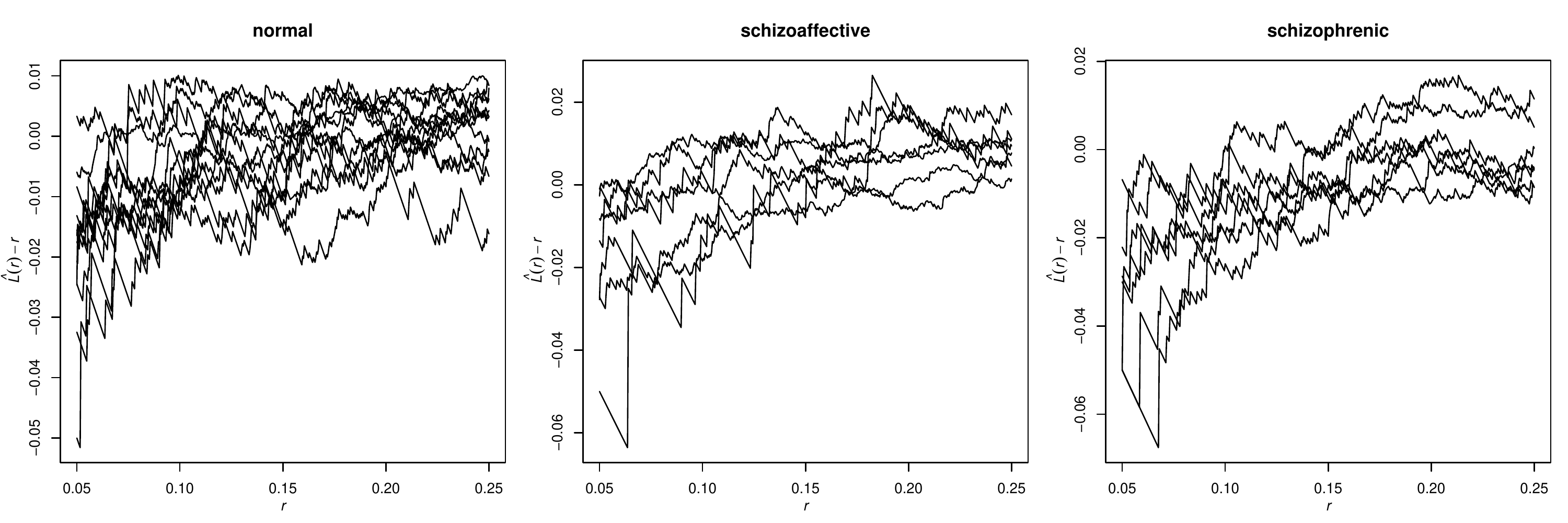} }
\captionof{figure}{\label{fig:groupL} The estimated centred $L$-functions in the three groups.}
\end{figure}
\end{center}

As a test function we chose the estimator of the $L$ function with the isotropic correction. Figure \ref{fig:groupL} shows these estimated $L$-functions in the three groups.

\subsection{Functional ANOVA}\label{sec:fANOVA}
In many functional ANOVA papers, the fact that the functions are measured only in a finite set of distances $r$ is utilized. Because of this discretization, it is possible to apply any ANOVA analysis for every $r$-value separately and obtain $K$ dependent values of an ANOVA statistic, where $K$ stands for the number of distances $r$. The statistic can be for example $F$-value of the $F$-test, log likelihood, BIC or a statistic of a nonparametric test. Then the test vector used in the rank test is 
$$\TT = (F(r_1), F(r_2), \ldots , F(r_K)),$$ 
where $F(r_i)$ stands for the chosen univariate statistic.
The simulations, which are necessary for applying the rank test, are produced by permuting the test functions.

As an illustration of this method we computed the one sided rank test for the $F$-statistics from 2499 permutations for the neuron data. The number of $r$ values was set to 500. The weighted ANOVA was performed in order to deal with unequal group variances of $L$-functions, which arises from the unequal mean numbers of points in the point patterns in different groups. The Kruskal-Wallis test with the $\chi^2$-statistic could be applied instead. 

\begin{center}
\begin{figure}[htbp]
\centerline{\includegraphics[scale=0.55]{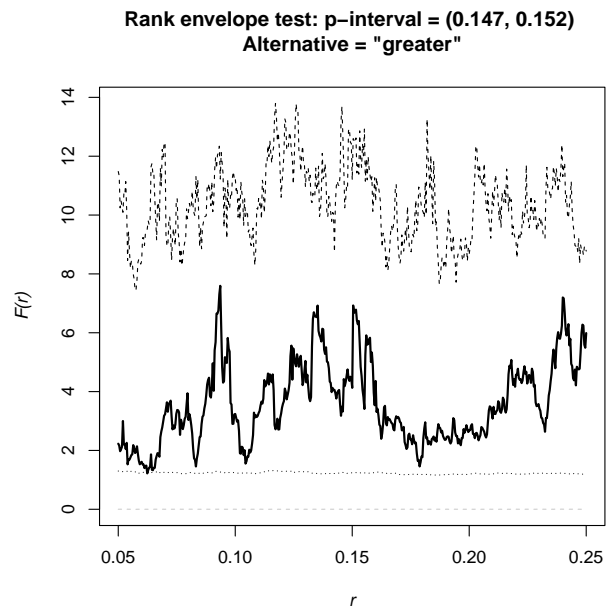}}
\captionof{figure}{\label{fig:fanova} Rank envelope test for comparison of 3 groups of $L$-functions via $F$-statistic of the weighted ANOVA done with $s=2499$ simulations. Solid line corresponds to the data $F$-function, the upper dashed line is the 95\% upper envelope, the lower dashed line is 0 corresponding to lower envelope and the dotted line corresponds to the average of $F$ functions from the permutations.}
\end{figure}
\end{center}

The resulted 95\% simultaneous upper envelope can be seen in Figure \ref{fig:fanova} showing no rejection in any distances $r$.

\subsection{One way group comparison}\label{sec:one-way-group-comparison}

Below we describe a new functional ANOVA procedure which is also based on our combined rank test and which is directly able to identify which groups are different and which distances are responsible for the possible rejection. All that is done at the common and exact significance level $\alpha$ guaranteed by the rank test.

Let us assume that we have $J$ groups which contain $n_1, \ldots , n_J$ test functions which are estimated from $n_1, \ldots , n_J$ point patterns and denote the test functions by $T_{ij}, i=1, \ldots, J, j=1, \ldots , n_j$. 
Assume that there exists not random functions $\mu(r)$ and $\mu_i(r)$ such that
$$ T_{ij} (r) =\mu(r) + \mu_i (r) + e_{ij}(r), i=1, \ldots, J, j=1, \ldots , n_j,$$
where $e_{ij}(r)$ are i.i.d. sample from a distribution $G(r)$ for every $r$. The only condition which $G(r)$ has to satisfy is that it has mean zero and finite variance. Thus we are dealing with completely nonparametric comparison of groups of functions.

We want to test the hypothesis $H_0$: 
$$H_0 : \mu_i(r) \equiv 0, i=1, \ldots , J.$$ 
This hypothesis can be clearly tested by the rank test, if the test vector is taken to consist of the average of test functions in the first group followed by the average of test functions in the second group, etc. 
We can shortly write that
$$\TT = (\overline{T}_1({\bf r}), \overline{T}_2({\bf r}), \ldots , \overline{T}_J({\bf r})),$$ 
where $\overline{T}_i({\bf r}) = (\overline{T}_i(r_1), \ldots , \overline{T}_i(r_K)$.
Thus, the length of the test vector becomes $J \times K$, where $K$ stands for the number of distances $r$. The simulations, which are necessary for applying rank test, are again produced by permuting the test functions $T_{ij}(r)$.  

The hypothesis $H_0$ is equivalent to the hypothesis 
$$H'_0 : \mu_i(r) - \mu_j(r) \equiv 0, i =1, \ldots , J-1, j =i, \ldots , J.$$ 
This hypothesis corresponds to the post-hoc test done usually after the ANOVA test is significant. 
However, this hypothesis can be directly tested by the combined rank test, if the test vector is taken to consist of differences of the group averages of test functions. We can shortly write that 
$$\TT' = (\overline{T}_1({\bf r}) - \overline{T}_2({\bf r}), \overline{T}_1({\bf r}) - \overline{T}_3({\bf r}), \ldots , \overline{T}_{J-1}({\bf r}) - \overline{T}_J({\bf r})).$$ Here the length of the test vector becomes $J (J-1) / 2 \times K$. 

Recall that both tests described above are done at one common significance level $\alpha$, which means that it is not necessary to perform the ANOVA test prior to the post-hoc test. Instead it is possible to apply only the post-hoc test obtaining an answer about the overall ANOVA test and also about the differences of groups.
Note that the two tests test the same hypothesis $H_0$ but the tests are not the same, they are sensitive to different departures from $H_0$. 

\subsubsection{Correction for an unequal variances for testing $H'_0$}

The two above procedure can be applied if the variances are equal across the groups of functions. 
To deal with different variances of group means of test functions in the different permutations, we rescale them to unit variance. Then the test vector becomes
\begin{equation}\label{TT_1prime}
\TT'_1 = \left(\frac{\overline{T}_1(r) - \overline{T_2}(r)}{\sqrt{\text{Var}(\overline{T}_1(r))+\text{Var}(\overline{T}_2(r))}},  \ldots , \frac{\overline{T}_{J-1}(r) - \overline{T_J}(r)}{\sqrt{\text{Var}(\overline{T}_{J-1}(r))+\text{Var}(\overline{T}_J(r))}}\right).
\end{equation}
In practice, $\text{Var}(\overline{T}_1(r))$ must be estimated for each $r$. 
For small samples, the sample variance estimator can have big variance, which may influence the procedure. 
The variance can be smoothed by applying moving average to the estimated variance with a chosen window size $b$ and replacing the sample variance in \eqref{TT_1prime} by its moving average analogue,
{\small$$\TT'_2 = \left(\frac{\overline{T}_1(r) - \overline{T_2}(r)}{\sqrt{\text{MA}_b(\text{Var}(\overline{T}_1(r)))+\text{MA}_b(\text{Var}(\overline{T}_2(r)))}},  \ldots , \frac{\overline{T}_{J-1}(r) - \overline{T_J}(r)}{\sqrt{\text{MA}_b(\text{Var}(\overline{T}_{J-1}(r)))+\text{MA}_b(\text{Var}(\overline{T}_J(r)))}}\right).$$}


\subsubsection{Weighted average}

Finally, instead of the basic average of functions $\overline T$ it is possible to use weighted average of functions in order to decrease the variance of group average and to give more weight to those functions that are more trustworthy. For example in the case of point pattern comparison, it is possible to follow \citet{DiggleEtal2000} and apply the weighted average on $L$-functions, where the weigths correspond to the number of points in the point pattern. 
In fact, the variance of the estimated $L$-function behaves approximately as $1/m_{ij}$, where $m_{ij}$ is the number of points in $ij$ point pattern. Thus, such weighted average can decrease the group variability of $L$-functions and improve the procedure.  
Thus for point patterns, the weighted average is defined as
$$\overline{L}_i(r)=\sum_{j}^{n_i} \frac{m_{ij}}{m_i}L_{ij}(r),$$ 
where $m_i=\sum_{j=1}^{n_i} m_{ij}$. The variance of weighted average, 
$$\text{var}(\overline{L}_i(r))=\sum_{j}^{n_i} \frac{m_{ij}^2}{m_i^2}\text{var}(L_{ij}(r)),$$
has to be then used in the $\TT'_2$. The variance $\text{Var}(L_{ij}(r))$ can be estimated as above.

\begin{center}
\begin{figure}[htbp]
\centerline{\includegraphics[scale=0.45]{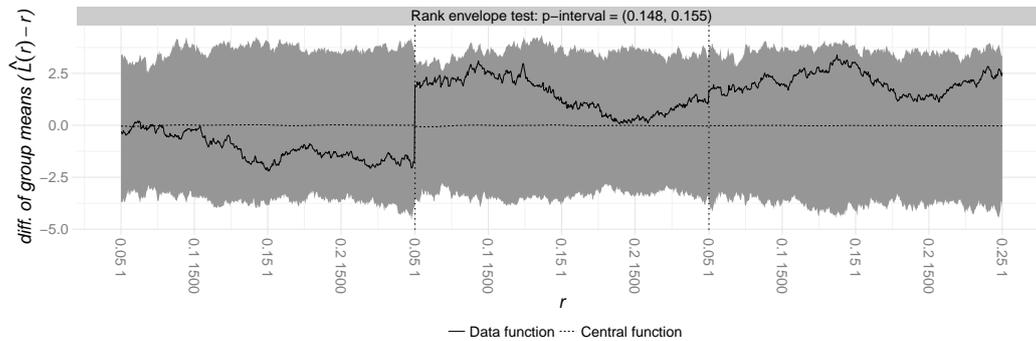}}
\captionof{figure}{\label{fig:3groupstukey} Rank envelope test for comparison of 3 groups of $L$-functions via difference of group weighted averages done with $s=7500$ simulations. The left subplot corresponds to the difference between the first and second group, the middle subplot corresponds to the difference between the first and third group and the right subplot corresponds to the difference between the second and third group. The grey area represents the 95\% global envelope.}
\end{figure}
\end{center}

Figure \ref{fig:3groupstukey} shows the result of the comparison of 3 groups of point patterns via difference of group weighted averages when $\TT'_2$ was used as test vector. The number of $r$ values was set to 500 and the window size of the moving average was set to 75. Each subplot shows the comparison of 2 groups. The test statistic being positive corresponds to the situation that the first group is more clustered than the second group in the comparison. Our result shows no differences between groups similarly as in the originally study and as the functional ANOVA test shown in Figure \ref{fig:fanova}. 

\begin{center}
\begin{figure}[htbp]
\centerline{\includegraphics[scale=0.45]{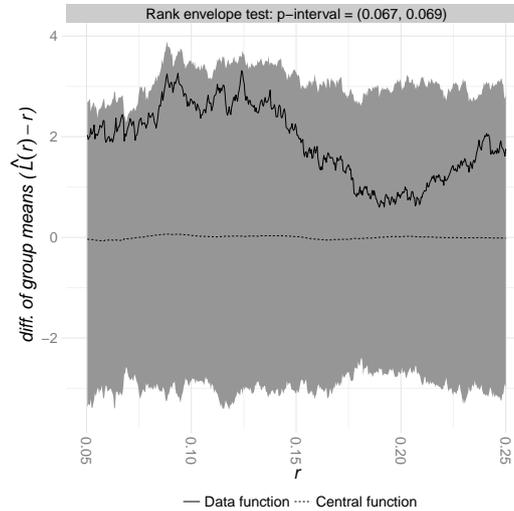}}
\captionof{figure}{\label{fig:2groupstukey} Rank envelope test for comparison of 2 groups of $L$-functions via difference of group averages done with $s=2500$ simulations. The plot corresponds to the difference between the joined first and second group and the third group. The grey area represents the 95\% global envelope.}
\end{figure}
\end{center}

We can observe that the first and second group are rather similar (the first subplot). Therefore we join the first and second group as was done in the original study. The result of the comparison of first and second group with third group is shown in Figure \ref{fig:2groupstukey}. 

Here we also do not observe significant difference between the first and second group with respect to the third group at the significance level 0.05.  Note that the original study reported a $p$-value of similar size, but \citet{Hahn2012} found out that the original method was liberal. 

\subsubsection{Correction for an unequal variances for testing $H_0$}

If many groups should be compared the above test based on group differences will consist of many subtests and the power can be small. In such a case it is possible to return to the test of hypothesis $H_0$ which consists of fewer subtests. To account for unequal variances in this test it is possible to set the test vector of the rank test as: 
{\small
$$\TT_2 = \left(\frac{\overline{T}_1(r) - \overline{T}_{-1}(r)}{\sqrt{\text{MA}_b(\text{Var}(\overline{T}_1(r)))+\text{MA}_b(\text{Var}(\overline{T}_{-1}(r)))}},  \ldots , \frac{\overline{T}_{J}(r) - \overline{T}_{-J}(r)}{\sqrt{\text{MA}_b(\text{Var}(\overline{T}_{J}(r)))+\text{MA}_b(\text{Var}(\overline{T}_{-J}(r)))}}\right),$$ }
where $\overline{T}_{-i}$ denotes the average of all test functions without the test function of the $i$-th group. Also the weighted average can be also applied in the same way as above.

\begin{center}
\begin{figure}[htbp]
\centerline{\includegraphics[scale=0.45]{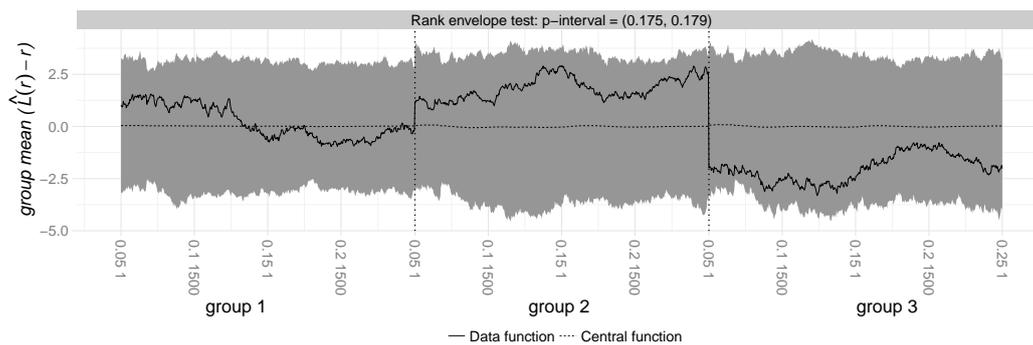}}
\captionof{figure}{\label{fig:3groupsmean} Rank envelope test for comparison of three groups of $L$-functions via difference of the averages of a group and the rest of the groups done with $s=7500$ simulations. The left subplot corresponds to the difference of the first group with respect to the remaining groups, the middle subplot corresponds to the second group and the right subplot corresponds to the third group. The grey area represents the 95\% global envelope.}
\end{figure}
\end{center}

Figure \ref{fig:3groupsmean} shows the comparison of 3 groups by means of $T_2$ with application of weighted average. Each subplot shows the comparison of a group with respect to the rest of groups. The test statistic being positive corresponds to the situation that the group is more clustered than the rest of groups. Our result again shows no differences between groups.

%

\section{Test for dependence of components in multi-type point processes} \label{sec:multitype}

The random superposition hypothesis for a bivariate point process says that
the point process results from the union of two independent components.
Testing of such independence between two types of points 
is typically based on the bivariate $L_{12}(r)$ function.
Simulations under this hypothesis are obtained by keeping the points of type 1 fixed,
and shifting the points of type 2 with periodic boundary conditions \citep[see e.g.][]{IllianEtal2008}.

For testing independence between $n>2$ types of points, to best of our knowledge,
there are no ``multivariate'' $L$ functions available.
Thus, a typical way to test the independence of $n>2$ sub-point patterns
is by going through all the pairs of types and performing consequently $n(n-1)/2$ tests,
where the test of independence of points of type $i$ and $j$ is
based on a bivariate $L_{ij}$ function ($i,j \in \{1, \dots, n\}$).
By means of the rank test, we can combine these tests to one test, i.e.\
to a test for the random superposition of $n>2$ components.

\begin{figure*}[h!]
\centering
\includegraphics[width=1\textwidth]{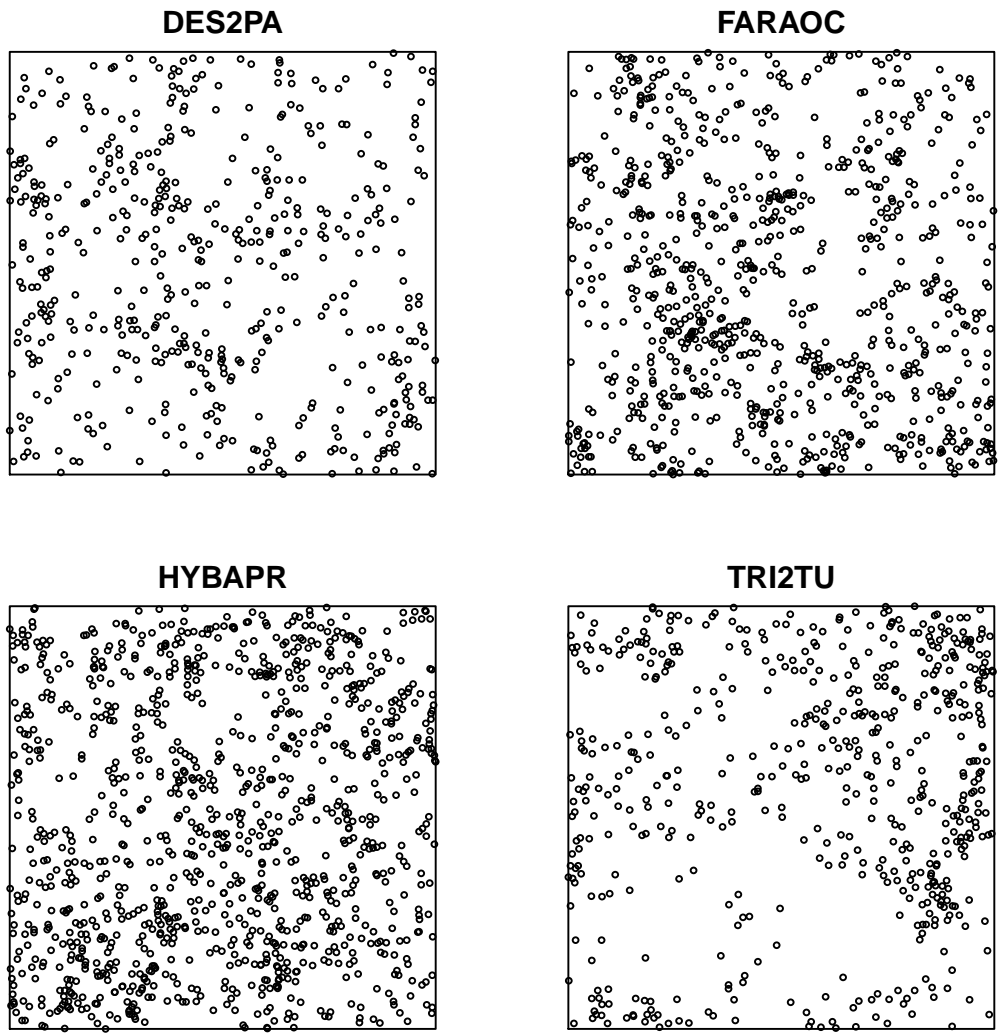}
\captionof{figure}{\label{Fig_rainforest}Four rainforest species in an area of size $100$ m $\times 100$ m.
DES2PA: Desmopsis panamensis; FARAOC: Faramea occidentalis, HYBAPR: Hybanthus prunifolius; TRI2TU: Trichilia tuberculata.}
\end{figure*}

We demonstrate this test for the patterns of the four richest tree species
in an area of size $100$ m $\times 100$ m in a tropical rainforest at Barro Colorado Island, Panama,
see Figure \ref{Fig_rainforest}.
The data origins from a 50 ha Forest Dynamics Plot in 2005, see \citet{HubbellEtal2005},
\citet{Condit1998} and \citet{HubbellEtal1999}.

Thus, we would like to test whether there are any small scale interactions between these species.
For this purpose, we used the $L_{ij}(r)$ functions on $I=[1, 25]$ and the multiple rank test
with six sub-tests, $(i,j) \in \{ (1,2), (1,3), (1,4), (2,3), (2,4), (3,4) \}$.
For computational reasons, we used only $s=4999$ simulations and calculated the extreme rank count $p$-value.
The obtained joint $p$-value of the test is $0.3416$,
while the obtained $p$-interval is $(0.3388, 0.3516)$, thus giving the same test result ``no rejection''. 
Thus, according to this test, we have no evidence against the random superposition of the patterns of the four individual species.

\section{Discussion}
In this paper, we have shown many possible applications of the rank (envelope) test for correction of multiple testing problem. The rank (envelope) test can be seen as a general solution to multiple Monte Carlo tests.
We have shown how the rank test can be used to perform a combined test for several univariate Monte Carlo tests, a combined test for pointwise Monte Carlo tests with a test function $T(r), r \in I$ and a combined test for several rank envelope tests performed with various test functions. The third case allows many interesting applications.

We considered the goodness-of-fit test with several test functions for a point pattern. In this case we have shown that using more test functions decreases the power of the combined test just a little in comparison to using the "best" function. Therefore a whole range of test functions can be used so that the test is sensitive to "all" possible deviations from the null model. Our simulation study also shows that using a powerless test function in the set of test functions does not decrease the power of the combined test much.

We also employed the goodness-of-fit test for several point patterns simultaneously. This problem may also be solved by computing the average test function over the point patterns and by comparing it with its simulated counterparts. Since the combined rank test compares each point pattern with the null model individually (but simultaneously at the global level $\alpha$), we believe it can lead to higher power than the other approach if the point patterns deviate from the null model in different ways.

We considered comparison of several groups of point patterns. Since a test function is used in the test instead of a point pattern, this test can be applied to any functional data and it can be seen as a functional ANOVA. Usually the test function is summarized into one number and  then the classical ANOVA is applied or a bootstrap method is applied. In our suggested approaches the whole test function is used and graphical interpretation shows which distances of the test function lead to a potential rejection. Additionally, in the second suggested approach, it is also seen which group leads to the possible rejection. This second approach can be seen as a post hoc test, which is performed at the exact significant level. This can be seen as an advantage also with respect to the classical ANOVA, because after an ANOVA test one has to perform a further post hoc test in order to find out between which groups there is a difference. Such post hoc tests are a bit conservative with respect to original ANOVA test, whereas in our approach the post hoc test is performed on the exact significance level. (The null hypothesis is simple in this case.)

Finally, we have applied the combined rank envelope test to the test of dependence of components in a multi-type point pattern with more than two components. Since performing the rank envelope test with many subtest (for many components) needs many simulations, we showed here the possibility of using a lower number of simulation together with our solution for ties (the extreme rank count ordering).

The aim of this paper was to show possible applications of the combined rank envelope test and its advantages. We are sure that this is not an exhausting list of the applications. There are further applications, e.g., in the fields of functional data analysis, geostatistics or random set theory.

\section*{Acknowledgements}
Mrkvi\v cka has been financially supported by the Grant Agency of Czech Republic (Projects No.\ P201/10/0472) and Mari Myllym{\"a}ki by the Academy of Finland (project number 250860). Hahn's research has been supported by  the Centre for Stochastic Geometry and
Advanced Bioimaging, funded by the Villum Foundation. The authors thank William R.Kennedy, Gwen Wendelschafer-Crabb and Ioanna G. Panoutsopoulou for providing the ENF data and Torsten Mattfeldt for providing the intramembranous particle data. 
The rainforest data set origins from the Forest
Dynamics Plot of Barro Colorado Island, which is made possible through the generous support
of the U.S. National Science Foundation, the John D. and Catherine T. MacArthur Foundation,
the Smithsonian Tropical Research Institute and through the hard work of over 100 people from
10 countries over the past two decades. The BCI Forest Dynamics Plot is part of the Center
for Tropical Forest Science, a global network of large-scale demographic tree plots.

\bibliography{Maris_bibfile} 
\end{multicols}
\end{document}